# The Apache Point Observatory Catalog of Optical Diffuse Interstellar Bands

Haoyu Fan

Key Laboratory of Optical Astronomy, National Astronomical Observatories, Chinese Academy of Sciences, Datun Road 20A, Beijing, 100012, China; University of Chinese Academy of Sciences, Beijing, 100049, China; hyfan@bao.ac.cn

L. M. Hobbs

The University of Chicago, Yerkes Observatory, 373 W Geneva St, Williams Bay, WI, 53191, USA

Julie A. Dahlstrom

Department of Physics and Astronomy, Carthage College, 2001 Alford Park Drive, Straz Center 94, Kenosha, WI, 53140, USA

Daniel E. Welty

Space Telescope Science Institute, 3700 San Martin Drive, Baltimore, MD, 21218, USA

Donald G. York

Department of Astronomy and Astrophysics, The University of Chicago, 5640 S. Ellis Ave, ERC 577, Chicago, IL, 60615, USA; The Enrico Fermi Institute, University of Chicago, IL, USA; don@oddjob.uchicago.edu; 773 702 8930

Brian Rachford

Embry-Riddle Aeronautical University, 3700 Willow Creek Road, Prescott, AZ, 86303, USA

Theodore P. Snow

University of Colorado, 2000 Colorado Ave, Duane Physics Building, Rm. E226, Boulder, CO, 80309, USA

Paule Sonnentrucker

Space Telescope Science Institute, 3700 San Martin Drive, Baltimore, MD, 21218, USA; European Space Agency

Noah Baskes

Department of Astronomy and Astrophysics, The University of Chicago, 5640 S. Ellis Ave, ERC 577, Chicago, IL, 60615, USA

Gang Zhao

Key Laboratory of Optical Astronomy, National Astronomical Observatories, Chinese Academy of Sciences, Datun Road 20A, Beijing, 100012, China; University of Chinese Academy of Sciences, Beijing, 100049, China; gzhao@nao.cas.cn

**ABSTRACT**: Aiming for a new and more comprehensive DIB catalog between 4,000 and 9,000Å, we revisited the Atlas Catalog based on the observations of HD 183143 and HD 204827 (Hobbs et al. 2008 and 2009). Twenty-five medium-to-highly reddened sight lines were selected, sampling a variety of spectral types of the background star and the interstellar environments. The median SNR of these spectra is ~ 1,300 around 6,400Å. Compared to the Atlas Catalog, 22 new DIBs were found, and the boundaries of 27 (sets) of DIBs were adjusted, resulting in an updated catalog containing 559 DIBs that we refer to as the *Apache Point Observatory Catalog of Optical Diffuse Interstellar Bands*. Measurements were then made based on this catalog. We found our survey most sensitive between 5,500 and 7,000Å due largely to the local SNR of the spectra, the relative absence of interfering stellar lines, and the weakness of telluric residuals. For our data sample, the number of DIBs detected in a given sight line is mostly dependent on $E_{B-V}$ and less on the spectral type of the background star. Some dependence on $f_{H2}$ is observed, but less well-determined due to the limited size of the data sample. The variations of the wavelengths of each DIB in different sight lines are generally larger than those of the known interstellar lines $CH^+$, CH, and K I. Those variations could be due to the inherent error in the measurement, or to differences in the velocity components among sight lines.

# 1. INTRODUCTION

As a long-standing mystery, the unidentified diffuse interstellar bands, or DIBs, play important roles in the interstellar medium network of constituents. The references to DIBs can be dated back to the year of 1922 (Heger 1922; Herbig 1995; McCall & Griffin 2011), almost a century ago, and there are hints of the presence of the strong and broad $\lambda$4428.8 DIB in an even earlier archive plate of HD 80077 (Code 1958; Oka & McCall 2011). Shortly after the discovery of DIBs, these absorption lines were identified to originate in the interstellar medium (e.g., Merrill 1934, 1936; Swings 1937; Swings & Rosenfeld 1937; Douglas & Herzberg 1941). Molecular carriers are then strongly suggested as the DIBs are much broader than lines of atoms or diatomic molecules in the same sight lines, and as substructures within some DIBs are observed even in "simple" sight lines (e.g. Sarre et al. 1995; Cami et al. 1997; Kerr et al. 1998; Galazutdinov et al. 2008). Despite the recent discussions regarding the $C_{60}^+$ ions as the carrier of five near IR DIBs (Campbell et al. 2015; Walker et al. 2015; though see Galazutdinov et al. 2017), the identification for the carriers of the vast majority of DIBs is still uncertain (Herbig 1995 for a review).

Heavily reddened early type stars most often serve as background sources for DIB detections (e.g. Jenniskens & Desert 1994; Galazutdinov et al. 2000; Tuairisg et al. 2000; Weselak et al. 2000). Thanks to the development of better instrumentation and improvement in the quality of the spectra, more and more DIBs are being found in the optical region as well as in the infrared (Geballe et al. 2011; Cox et al. 2014; Hamano et al. 2016; Elyajouri et al. 2017). Based on the observations toward the two sight lines HD 183143 (B7Iae, $E_{B-V}$ = 1.27 mag) and HD 204827 (O9.5V + B0.5III, $E_{B-V}$ = 1.11 mag), Hobbs et al. (2008 and 2009) created two independent atlases of optical DIBs that are the basis for this merged catalog that includes new observational material. These two atlases were produced independently and the two stars will be referred to in this paper as the atlas stars.

On the other hand, the kinds of DIBs presented in the two atlas sight lines are different from each other. An unpublished comparison was carried out in 2010 by L.M.H. of the 380 DIBs detected toward HD 204827 (Hobbs et al 2008) and the 414 DIBs detected toward HD 183143 (Hobbs et al 2009). The comparison was based effectively on the DIBs' central wavelengths, after the differences between the radial velocities of the clouds along the two sight lines were taken into account, as indicated by the respective K I 7699Å line profiles. Among a total of 545 distinct DIBs, 249 were common to both sightlines, while 131 DIBs were detected toward HD 204827 only, along with 165 toward HD 183143 only (specific numbers depend on the adopted detection limits). The difference between the DIBs present in the two sight lines might be due to the very different interstellar environments, as indicated by the presence of the strong interstellar lines of the $C_2$ and $C_3$ molecules in the sight line of HD 204827, which are not seen towards HD 183143 (Oka et al. 2003). A question then arises: does the difference in the entries of the two atlas stars encompass all DIBs, or will more such differences be found when cataloguing additional sight lines exhibiting physical conditions somewhat different from HD 183143 and HD 204827?

We hope to create a more comprehensive DIB catalog on the basis of the findings by Hobbs et al., namely, to see if new DIBs exist in other sight lines, and to see if the profiles and wavelengths of the known DIBs in the atlas stars are consistent when observed in the other sight lines. To serve this purpose, high signal to noise ratio (S/N ratio), moderately high resolution spectra of twenty-three sight lines, as well as the two atlas stars used in Hobbs et al. (2008 and 2009), were selected from our database. These sight lines have $E_{B-V}$ ranging between 0.31 and 3.31 mag and cover a diversity of the ISM conditions. Their background stars also cover a variety of spectral types. We looked for new DIBs in these spectra, while revisiting and checking the robustness of the known DIBs (in terms of repeatable detections) with a different, semi-automated method of measurement.

This work is organized as follows. Sections 2 and 3 describe the target sight lines selected for this work and general data reduction process. We report our attempts to refine the current list of DIBs in Section 4 and the resulting *Apache Point Observatory Catalog of Optical Diffuse Interstellar Bands* (APO Catalog of DIBs, for short) is given in Section 5. After that, Section 6 compares the detection percentages and profile properties of the DIBs to different parameters. And finally, the conclusions reached in this work are briefly summarized in Section 7.

## 2. Observation and Data Reduction

All spectra used in this work were obtained with the 3.5m telescope and the ARC echelle spectrograph (ARCES; see Wang et al. 2003) at the Apache Point Observatory (APO). The wavelength coverage of ARCES is between 3,500Å and 11,000Å, and the resolving power is about 38,000, corresponding to a 9 km s$^{-1}$ velocity resolution (2 pixel). The typical single exposure time was 20 minutes for stars fainter than V ~ 7.0 magnitude, to minimize the impact of cosmic rays. High signal to noise ratio (S/N ratio) for fainter stars was achieved by combining multiple exposures (up to 35) of the same target, and a typical nominal S/N ratio of ~1,000 per resolution element around 6,400Å was achieved for our archive of some 400 stars.

We note that the spectra used in this work are part of the database of the DIB survey project we have been undertaking since 1999. Thus, some of these spectra have also been used in other papers of this project, e.g. the series-ID papers "Studies of Diffuse Interstellar Bands" (Thorburn et al. 2003, Hobbs et al. 2008, Hobbs et al. 2009, McCall et al. 2010, and Friedman et al. 2011), and the related papers (McCall et al. 2001, Dahlstrom et al. 2013, Welty et al. 2014, and Fan et al. 2017). Readers can refer to these works for more detailed data reduction processes and related discussions. We only include a brief description in this work.

The basic data reduction of the raw spectra was done by J. D. and described in Thorburn et al. (2003). To correct telluric lines, a model spectrum is fitted based on air mass and humidity and then removed from the original astronomical spectrum. The isolated, weak telluric lines are usually well corrected. It is easy to see when the corrections in a given star at a given wavelength are poor and we do not try to measure DIBs at such locations, even if they can be seen to be present. The wavelength scale of all the program and comparison spectra in this work are set to the interstellar frame of reference by setting the apparently strongest component of the interstellar K I line at laboratory wavelength 7698.9645Å (Morton 2003) to zero velocity.

## 3. Program Sight Lines

In total, twenty-five spectra from medium to highly reddened early type stars were selected, and will be referred to as program stars in this work. The basic information of the program stars and the ISM properties along the sight lines are provided in Table 1. Each program star is paired with a low reddened star with the same or similar spectral type, luminosity class and ideally with low projected rotational velocity (*vsini*). These low reddened stars will be noted as comparison stars, and their information is also provided in Table 1. The low reddening of the comparison stars ensures minimal observed strength of interstellar lines (including DIBs), if present. Their spectra provide clean references for potential stellar line contaminations in the program stars, although the relative strengths of some stellar lines may differ even for stars of the same nominal spectral type.

In this section, we include summaries of the general properties of the program sight lines chosen for the search of DIBs, as well as particular sight lines of special interest. Although the completeness for the list of DIBs cannot be tested, we emphasize that the great diversity of the target sight lines helps to promote the completeness of the DIB catalog reported later in the paper.

3.1 Program Stars and Their Spectra

To distinguish DIBs from stellar lines is crucial for all DIB studies. However, even if identified, the stellar lines still hinder the detection of DIBs at adjacent wavelengths. To minimize the impact from stellar lines, repetition in the spectral types of the program stars is (to the maximum degree) avoided in our data sample. Our program stars consist of six O-type stars, fourteen B-type stars and five A-type stars, and include stars known to be main sequence stars, giants, and supergiants. Twelve out of the twenty-five program stars show up as spectroscopic binaries in spectra from our limited epochs of observations[1]. For these targets, the Doppler motions also help in identifying the interstellar lines from the stellar lines when vetting their spectra. The variety of spectral types reduces the chance of a DIB being repeatedly blocked by the same stellar line, or of a false identification being triggered by misidentified stellar lines.

For a DIB of a given width, the equivalent width (EW) detection limit depends on the S/N ratio of the spectrum. We took the spectrum with the highest S/N ratio from our archive, when multiple choices of stars with similar spectral types, $E_{B-V}$, and the ISM conditions along the sight lines are available. The S/N ratio is mainly set by the magnitudes of our stars on the night of observation, the optical efficiency of the telescope and instrument system, and the weather conditions during observations. The average S/N ratio for the final, accumulated spectra of the twenty-five program stars near 6400Å is 1,750 and the median value is 1,300. These values are higher than the typical S/N ratio of ~1,000 in our larger database.

---

[1] These stars are HD 23180, HD 24534, HD 166734, HD 168625, HD 175156, HD 190603, HD 194279, HD 204827, HD 206267, HD 223385, HD 281159, and BD+40 4220 (also known as VI Cyg 5 and Cyg OB2 5). See also Table 1 of this work.

The distances of the program stars affect the volume of space sampled for DIBs. Distances accurate to better than 20% are available from GAIA for 18 of our stars (Gaia Collaboration et al. 2018). Six additional stars have photometric distances (Neckel et al. 1980). For these 24 stars, the average distance is 1000pc and the median distance is 600 pc. The nearest star is HD147084 (omicron Sco) at 126pc and the most distant is HD 190603 at 3948 pc. The only star we could not find a reliable distance for is BD +31 640. The program stars are mostly distributed in the plane of the Milky Way, with a few of the closer stars as far as 20 degrees off the plane. A few of the program stars are in tight groups with possible or confirmed regional differences: Scorpius (3 stars), Cygnus (3 stars), and Perseus (IC348, 5 stars; Sonnentrucker et al. 1999).

3.2 The ISM along the Program Sight Lines

The reddening is one of our prime considerations during the selection of the program stars. The sight line must be sufficiently reddened to promote the detections of DIBs, especially for the weaker ones. Heavily reddened sight lines are also more likely to contain clouds of different types where new DIBs might be found (Bailey et al. 2015). The reddening of the program sight lines in this work range from 0.31 mag (HD 23180/omicron Per and HD 175156) to 3.31 mag (VI Cyg 12/Cyg OB2 12); the average and median values are, respectively, 1.02 and 0.67 mag. These reddenings are comparable to the two atlas sight lines used in Hobbs et al. (2008 and 2009, $E_{B-V}$ = 1.11 and 1.27 mag respectively, which are included and re-measured in this work). There are a few sight lines with $E_{B-V}$ less than 0.5 mag, which is a necessary compromise to other considerations related to the ultimate comprehensiveness of the Catalog, as described below.

The sample includes 17 stars for which the UV extinction curves have been published using the techniques of Fitzpatrick & Massa (1990 and 2007). Four of these sight lines (HD 23180, HD 23512, HD 24534, and HD 28482) have unusually steep far-UV rises along with broad 2175Å bumps. Another two stars (HD 175156 and HD 204827) have steep far-UV rises but with weak bumps of normal width, while HD 147889 has steep far-UV extinction and normal bump. Four sight lines have shallow far-UV extinction combined with weak 2175A bumps (HD 37061, HD 37093, HD 148579, and HD 190603). The remaining six sight lines (HD 24912, HD 166734, HD 168625, HD 183143, HD 206267, and HD281159) have normal far-UV extinction.

Plenty of previous studies have shown that, besides the reddening, other environmental conditions along the sight line also affect the strengths and profile of some DIBs. Such examples include the lambda-shaped behavior of $W$(DIBs)/$E_{B-V}$ when plotted against the fraction of molecular hydrogen[2] ($f_{H2}$; Fan et al. 2017 and references within); the effect of radiative pumping as seen towards Herschel 36 (Oka et al. 2013); and the various indications found in Galazutdinov et al. (2015). The $f_{H2}$ value can be used as a general indicator of the average ISM along the sight line (see the discussions in Fan et al. 2017). For the twenty-five program sight lines of this work, it has a range of 0.02 (HD 37061) to > 0.78 (Cernis 52), and its average and median values are 0.43 and 0.44 respectively.

The low abundances of certain atomic and molecular species and anomalously weak DIBs are often attributed to the presence of strong radiation fields (e.g. Savage et al. 1977; Herbig 1993; Welty & Hobbs 2001) because these species are either ionized or dissociated via photon processes. On the other hand, when dense cloud components are involved in the sight line, the strengths of some DIBs are also found to be depressed (e.g., Wampler 1966; Adamson et al. 1991; Fan et al. 2017 and references therein). Such behavior might be related to the competition between radiation field and shielding effect from the ISM cloud, and subsequent ionization via photon processes (see e.g. Jenniskens et al. 1994; Sonnentrucker et al. 1997), although other mechanisms such as depletion and hydrogenation/de-hydrogenation may be involved as well (Vuong & Foing 2000).

If the destruction of DIB carriers is governed by photon processes, their fragments might be found in sight lines containing intense radiation fields where photon dissociation takes place efficiently. Meanwhile, the precursors of DIB carriers might be found in dense cloud components. Additional DIB-like features may be found in both cases, and it would be interesting to see if such features lie in the optical region as new DIBs (Cami et al. 1997), or if certain groups of DIBs show up uniquely under certain sets of interstellar conditions as was seen for the $C_2$ DIBs (Thorburn et al. 2003; Elyajouri et al. 2018).

The mean intensity of the interstellar radiation field (ISRF) at 1300Å at the mid-point of each of our sight lines was calculated based on a new evaluation by Adolf Witt and Ethan Polster (private communication). The calculation uses Gaia parallaxes (Gaia Collaboration et al. 2018) of the background stars, and places them in the integrated field. For the twenty-five program sight lines, the mean ISRF at 1300Å averages $3.7 * 10^{-6}$ ergs/cm$^2$-sec-Å, and the median value is $1.5 * 10^{-6}$ ergs/cm$^2$-sec-Å. Five of the program stars (HD 23512, HD 37903, HD 147084, HD 147889, and HD 148579) stand out as having the highest radiation fields, with an average value of $13 * 10^{-6}$ ergs/cm$^2$-sec-Å, whereas the average value for the remaining 20 stars is $1.3 * 10^{-6}$ ergs/cm$^2$-sec-Å. These numbers are meant to be indicative of sight lines that could be subject to strong ISRF, rather than for direct usage since the distance of the IS cloud(s) is not considered. But their variety here does indicate the variation of ISRF along our program sight lines, which is the point of the discussion here.

---

[2] $f_{H2} = 2*N(H_2)/(N(H) + 2 * N(H_2))$

The order of magnitude range exhibited by the W(5780)/W(5797) ratio in different sight lines, known as the sigma-zeta effect (named after the early examples σ Sco and ζ Oph; Sneden et al. 1991; Kre łowski et al. 1992), is often associated with the strength of the local UV radiation field. In this picture, the carrier of the 5797 DIB is more easily destroyed by UV radiation than the carrier of the 5780 DIB, so that small W(5780)/W(5797) ratios (1 - 2) are found for sight lines with weaker UV fields, while sight lines with stronger UV fields have larger ratios (up to ~10; e.g. Vos et al. 2011; Fan et al. 2017). In our data sample, the W(5780)/W(5797) ratio ranges between 1.1 and 8.9, with a median value of 3.1. Both the overall range and the typical value for these twenty-five sight lines thus are similar to those seen in the larger data sample of more than 180 sight lines used in Fan et al. (2017)[3], as well as those found in other studies of DIBs in the Galactic ISM -- suggesting that these twenty-five sight lines sample a range in UV field strength.

While we know little in detail about how the UV ISRF affects the DIBs, we do know how it affects some identified molecules. It is thus important to find robust patterns or correlations between those known molecules and DIBs, to better connect the DIBs to known physical quantities, such as the radiation field or the volume density of hydrogen (e.g. the $C_2$ DIBs and the C molecules, see Thorburn et al. 2003). Our sample is heavily selected toward sight lines with detections of diatomic (CN, CH) or triatomic molecules ($C_3$, $H_3^+$). Randomly chosen sight lines would largely miss these regions, but those with the highest column densities where DIBs might be easiest to detect, and those involving concentrations (highest volume densities) where different kinds of DIBs may be found, are prime targets of this survey. An extensive data collection is maintained by one of us (D.E.W.). These data are obtained from measurement of archival spectra and searches of the literature. Detection of, or sensitive limits for, CH$^+$, CH , $C_2$ and CN in common are available for 21 stars, and two more stars missing only the measurement of $C_2$. Of these stars, half have sensitive detections of $H_3^+$ and/or $C_3$. In addition, nine stars in our sample have sensitive detections or limits of CO and six have measurements on OH. Among the 21 sight lines noted above, the column density varies over a factor of 29 for $C_2$, 50 for CH, 93 for $CH^+$, and 251 for CN. The range of variation is much less for $H_3^+$ and OH, but much higher for CO, though there are fewer measurements in all three cases for observational reasons. The ratios of $N(CH^+)/N(CH)$ and $N(CN)/N(CH)$ both spread over a factor of ~35 and suggest great variety of the ISM environments found along the program sight lines of this work (Federman et al. 1994; Godard et al. 2014).

---

[3] Note that while some of the spectral data used here are the same as in Fan et al. (2017), the ratios reported are based on the new measurements made for this project. Any difference in the measured EWs and strength ratios are within the mutual uncertainties of measurement.

Each of the molecules exists under different optimal conditions, which can in principle be discerned by inter-comparing observations. For instance, $H_3^+$ is sensitive to the low energy cosmic ray flux (Indriolo et al. 2007), CN is thought to exist under high density conditions (Federman et al. 1994), and $CH^+$, while of uncertain environment, evidently forms via non-equilibrium processes (Godard et al. 2012).

Assuming that the denser cloud conditions in the interstellar medium under which these molecules exist may have different morphologies (sheets, filaments, approximate spheres, etc.), the conditions from the outside to the inside of an identifiable region are presumably continuous and give rise to different molecules in different zones from lower to higher densities (Pan et al. 2005; Kos 2017). These different zones may have dimensions of AUs to pcs and each star probes all the density zones that happen to be present on a given sight line. The zones may be typically similar from one cloud situation to another, and the molecules that exist, regardless of the physical mechanism at play (history, formation, or destruction), presumably have repeatable patterns from one case to another. This is true provided the various typical zones are present, whereas HD 62542 is an example where the typical "zone" distribution has been altered (Cardelli et al. 1990; Snow et al. 2002; Welty et al. in preparation).

Similarly to known molecules, some of the DIB carriers may have related and repeatable patterns as for the $C_2$ DIBs and the $C_2$ molecules (Thorburn et al. 2003), and probing the many different zones indicated for diatomic or triatomic molecules may show up different DIB carriers (Welty et al. 2014; Fan et al. 2017). The wide range of molecular ratios indicated above over our particular sample of twenty-five stars provides a high probability that if related DIB carriers are present on these sight lines, they could show up, providing only that they are above our detection threshold. Fan et al. (2017) and reference within suggest that the strongest diffuse bands may, similarly, show a consistent spatial ordering in different clouds. Our sample may therefore include related zones of both DIBs and simple molecules and provide a fairly complete survey of such regions to promote the completeness of the DIB catalog.

3.3 Sight Lines of Particular Interest

Some of the targets in this study were selected due to their known unusual ISM conditions, as follows.

HD 37061 (NU Ori) is in the H II region M43, close to the Trapezium stars in the Orion region. The interstellar material is exposed to an intense radiation field from nearby early-type stars. This leads to a very small molecular fraction of hydrogen ($f_{H2}$ = 0.02), which can be used to characterize the average interstellar environment along the sight line. Many of the strong and well known DIBs are greatly weakened (compared to $E_{B-V}$) or even absent in the sight line, such as DIBs $\lambda\lambda$5780.6, 5797.1, 6196.0 and 6613.6 (Fan et al. 2017).

The sight line of HD 24534 (X Per) stands opposite to HD 37061 in terms of the radiation field and shielding of $H_2$. The translucent cloud known to have high molecular abundances (Mason et al. 1976; Lien 1984 a, b, and c; Federman & Lambert 1988) gives the sight line a rich interstellar molecular spectrum and a large value for $f_{H2}$ of 0.76, the second largest in our data sample (Table 1). The strength ratio W(5780)/W(5797) is 1.2 in this sight line (subject to the measuring technique applied, see discussions in Fan et al. 2017), which is one of the smallest ratios ever observed. The UV spectra available for this sight line have yielded measurements of total hydrogen density and the excitation temperatures of CO, $C_2$ and $H_2$ molecules (Sonnentrucker et al. 2007).

The W(5780)/W(5797) ratio generally decreases with increasing $f_{H2}$, but the sight line of HD 37903 is an extreme outlier in this regard, with a W(5780)/W(5797) ratio of 8.2 at $f_{H2}$ = 0.55 (Fan et al. 2017, Figure 5; subject to the measuring technique applied). This B1.5V star is embedded in the front part of the molecular cloud LDN 1630 (Witt et al. 1984). Vibrationally excited $H_2$ lines have been observed in this sight line (Meyer et al. 2001). However, the radiation input from this early B-type star is not as harsh as that to which the gas around the nearby stars in the Orion Nebula are exposed. This sight line is thus characterized by the unusual combination of an intense radiation field and a fairly high $f_{H2}$.

The young stars VI Cyg 5 and VI Cyg 12 belong to a massive OB association containing some 2,600 stars (Knödlseder 2000). It is behind the Great Cygnus Rift (Hanson 2003, Guarcello et al. 2012), and the dust in the dark lane results in very large extinction ($E_{B-V}$ = 1.99 mag for VI Cyg 5 and 3.31 mag for VI Cyg 12). The high column densities of the ISM species and large reddenings along these long sight lines make them perfect targets for detecting weak DIBs from dusty material (Chlewicki et al. 1986). The extinction along the sight lines is similar to typical interstellar material in the diffuse field (Whittet 2015), but the detections of various small molecules, such as $C_2$, CN and CO strongly favor the presence of dense cloud components. Mapping of the emissions from the CO molecules suggests that the molecular gas might have a clumpy distribution in this region (Scappini et al. 2002, Schneider et al. 2006). Thus, there can be complex density structures (Hamano et al. 2016).

The polycyclic-aromatic hydrocarbon (PAH) molecules have been thought by some investigators to be promising candidates for DIB carriers (e.g. Gredel et al. 2011), although no matches between the DIBs and PAH absorptions have been found. The simplest PAH molecule, the naphthalene cation ($C_8H_{10}^+$) was reported to be detected in the sight line of Cernis 52 (Iglesias-Groth et al. 2008; González et al. 2009), though Searles et al. (2011) and Galazutdinov et al. (2011) showed that the naphthalene lines, if present at the known lab wavelengths, would have too broad and shallow profiles and be hard to detect with echelle spectra with confidence (Hobbs et al. 2008, Sonnentrucker et al. 2018). This issue deserves further study with much better spectra than presently available. A microwave emitting cloud is at a similar distance of this star in the sight line (Cernis 1993). These anomalous microwave emissions (AMEs, Watson et al. 2005) can be caused by an enhanced presence of small spinning dust grains in the diffuse ISM (Draine & Lazarian 1998) but were suggested by Bernstein et al. (2015) to be associated with DIBs.

Our particular selection of twenty-five stars thus provides a set of search regions for DIBs that might exist in the interstellar medium, sampling a range of external effects such as varying UV and IR radiation field, properties of interstellar grains, cosmic radiation, stellar outflows, etc. It is to obtain the wide sample of different cloud morphologies and external conditions that we chose this sample and sought to obtain the highest signal to noise possible.

While this section points out the strengths of our selection of stars in probing a variety of unusual environmental conditions for a DIB search, the sample is nonetheless too limited to satisfy the need to explore all the places where DIBs might appear and also to satisfy the requirement that each DIB feature be detected in multiple sight lines (at least five, see Section 5). Some possible additional sites where new DIBs might occur are included in our survey by default. For instance, there are many more clouds without molecules than with molecules and the latter are probed by the lines of sight we have emphasized. Likewise, stellar outflows from hot stars and the gas in H II regions are covered. Meanwhile, comets and circumstellar disks are beyond the scope of this search, as are objects in the high halo of the Galaxy where the 21-cm high velocity clouds are found (Wakker & van Woerden 1997; Wakker et al. 2007). These are just three examples of specific easily identified additional locations which should be explored that could shed new light on the origin of the DIBs.

Table 1 Information of the twenty-five Sight Lines Selected for This Work

| Star Name[a] (HD/BD) | Identifier | Spectral Type | $E_{B-V}$ (mag) | vsini (km s$^{-1}$) | $f_{H2}$[b] | Det. Pct. - All[c]/Mea.[d] | Comp. Star (HD) | Spectral Type[e] | $E_{B-V}$[e] (mag) | Vsini[e] (km s$^{-1}$) | Comments |
|---|---|---|---|---|---|---|---|---|---|---|---|
| 20041 | | A0Ia | 0.72 | 29 | 0.42* | 62.8%/75.6% | 46300 | A0Ib | 0.01 | 14 | |
| BD+31 640 | Cernis 52 | A3V | 0.90 | 125 | >0.78* | 21.6%/23.8% | 107966 | A3V | 0.00 | 51 | Report of PAH, González et al. (2009); |
| 23180[f] | omi Per | B1III+B2V | 0.31 | 90 | 0.55 | 49.9%/52.3% | 44743 | B1II-III | 0.02 | 17 | Steep ext. curve; Broad 2175Å bump |
| 281159[f] | | B5V | 0.85 | 162 | 0.50* | 66.0%/66.6% | 16219 | B5V | 0.04 | 30 | |
| 23512 | | A0V | 0.36 | 140 | 0.62* | 21.1%/22.5% | 31647[e] | A1V | 0.01 | 105 | Steep ext. curve; Broad 2175Å bump |
| 24534[f] | X Per | O9.5pe | 0.59 | 200 | 0.76 | 40.4%/42.0% | 214680 | O9V | 0.11 | 35 | Translucent cloud; Steep ext. curve; Broad 2175Å bump |
| 24912 | xi Per | O7e | 0.33 | 213 | 0.38 | 53.7%/56.0% | 47839 | O7Ve | 0.07 | 70 | |
| 28482 | | B8III | 0.48 | 30 | 0.66* | 26.1%/31.0% | 4382 | B8III | 0.01 | 23 | Steep ext. curve; Broad 2175Å bump |
| 37061 | NU Ori | B1V | 0.52 | 160 | 0.02 | 25.8%/26.5% | 36959 | B1V | 0.03 | 5 | Intense radiation field; Flat ext. curve; Weak 2175Å bump |
| 37903 | | B1.5V | 0.35 | 200 | 0.53 | 29.3%/30.4% | 37018 | B1V | 0.07 | 20 | Anomalous 5780/5797 ratio; Flat ext. curve; Weak 2175Å bump |
| 43384 | 9 Gem | B3Ib | 0.58 | 35 | 0.44* | 65.3%/71.6% | 52089 | B2II | 0.01 | 25 | |
| 147084 | omi Sco | A5II | 0.73 | 19 | 0.59* | 19.5%/35.0% | 186377 | A5III | 0.04 | 15 | |
| 147889 | | B2V | 1.07 | 100 | 0.45 | 67.3%/69.5% | 42690 | B2V | 0.04 | <5 | Embedded and ionizing nearby cloud (Rawlings et al. 2013), Steep ext. curve; |
| 148579 | | B9V | 0.34 | 150 | 0.45* | 25.6%/27.1% | 201433 | B9V | 0.00 | 25 | Flat ext. curve; Weak 2175Å bump |
| 166734[f] | | O8e | 1.39 | 175 | 0.39* | 89.1%/90.9% | 47839 | O7Ve | 0.07 | 70 | |
| 168625[f] | | B8Ia | 1.48 | 50 | 0.33* | 62.8%/66.6% | 34085 | B8Iae | 0.00 | 40 | |
| 175156[f] | | B5II | 0.31 | 20 | 0.31* | 51.9%/58.7% | 34503 | B5III | 0.05 | 40 | Steep ext. curve; Weak 2175Å bump |
| 183143 | | B7Iae | 1.27 | 60 | 0.31* | 84.4%/92.5% | 63975 | B8II | 0.00 | 25 | Hobbs et al. (2009); |
| 190603[f] | | B1.5Iae | 0.72 | 35 | 0.16 | 57.4%/64.1% | 52089 | B2II | 0.01 | 25 | Flat ext. curve; Weak 2175Å bump |
| 194279[f] | | B2Iae | 1.20 | 70 | 0.30* | 61.4%/67.4% | 53138 | B3Iab | 0.05 | 35 | Multiple components but average condition (Cox et al. 2011); |
| BD+40 4220[f] | VI Cyg 5[g] | O7f | 1.99 | | 0.47* | 83.9%/86.4% | 47839 | O7Ve | 0.07 | 70 | |
| | VI Cyg 12[h] | B5Ie | 3.31 | 50 | >0.48* | 78.5%/82.1% | 164353 | B5Ib | 0.11 | 40 | Schulte's Star; |
| 204827[f] | | O9.5V+B0.5III | 1.11 | 105 | 0.67* | 87.8%/90.0% | 36959 | B1V | 0.03 | 5 | Hobbs et al. (2008), Steep ext. curve; Weak 2175Å bump |
| 206267[f] | | O6f | 0.53 | 155 | 0.42 | 66.9%/69.8% | 47839 | O7Ve | 0.07 | 70 | |
| 223385[f] | 6 Cas | A3Iae | 0.67 | 35 | 0.12* | 49.2%/63.1% | 197345 | A2Ia | 0.09 | 35 | |

a: Sorted by RA of the program star.

b: The mass fraction of molecular hydrogen to all neutral hydrogen along the sight line. If marked "*", computed by using $N$(CH) as a surrogate for

$N(H_2)$ and/or W(5780) for $N(H)$. See Fan et al. (2017) for details.

c: The number of detected DIBs in each sight line divided by 559, the number of DIBs reported in Table 2.

d: The number of detected DIBs divided by the total measuring attempts made in each sight line, which is 559 minus the number of rejections due to various reasons.

e: For the comparison star.

f: Spectroscopic binaries identified in our observations and/or Binary star DataBase (BDB, Kovaleva et al. 2015).

g: Often referred to as Cyg OB2 5.

h: Often referred to as Cyg OB2 12.

## 4. Data Analysis

In this work, our primary goal is to create a new and expanded catalog of DIBs. Unlike previously published catalogs of primarily narrow DIBs focused on only one or a few sight lines, twenty-five program sight lines with a variety of ISM conditions were selected in this work. We tried to cover all DIBs in the optical region with a sufficient number of detections in the spectral sample to be sure the claimed DIBs were real, and expected these DIBs to have similar profiles characterized by central wavelengths, widths, etc. in different sight lines. The consistency in the measurement of a given DIB is achieved by applying the same measuring technique in all sight lines, i.e. the continuum placement (e.g. using nearby region without any stellar or DIB absorption, or based on the profile of adjacent strong DIB), and a particular definition of the end points of the DIB region (e.g. when the DIB profile hits the continuum level, or the inflection point between two blended DIBs). We emphasize that this measuring technique provides reference to the measurement and no fixed wavelengths were assigned the continuum region or boundaries of any DIB. They were set according to the situation of each DIB in each star so as not to miss subtle but rare environmental effects that might introduce changes in normal profiles, central wavelengths or line widths. Readers can refer to the discussion on measuring the well-known DIB $\lambda$5797.1Å in the appendix of Fan et al. (2017).

Three major steps were taken to compile such a catalog,

1) Revisiting the identified DIBs in the Atlas Catalog with similar wavelengths in HD 183143 and HD 204827, to confirm them as common DIBs to the two atlas sight lines, or list them as candidates for further evaluation in the program spectra of this work;

2) Looking for new DIB candidates in the newly selected program spectra;

3) Measuring all confirmed DIBs in the twenty-five sight lines that will constitute our APO Catalog of DIBs, and checking the repeatability (wavelength and width) of each of these DIBs so as not to miss any subtle changes in the profiles, central wavelengths, or line widths caused by environmental effects.

### 4.1 Evaluating the Known DIBs

A comparison between the DIBs detected in the two atlas sight lines of HD 183143 and HD 204827 was performed in 2010 by L.M.H. to combine the results. This led to the unpublished Atlas Catalog used as the starting point for our extended DIB survey. This atlas is presented as column E (wavelength) and column H (FWHM) of Table 2. A key step in building this preliminary merged atlas was to identify the common DIBs in the two atlas stars, based primarily on agreement in the central wavelengths.

To begin the new evaluating of each DIB with respect to the expanded list of observed stars, we considered DIBs in the Atlas Catalog whose central wavelengths agreed to within 0.5Å and whose FWHM agreed to within 50% as referring to the same DIB. Cases of larger differences in wavelength and/or FWHM were examined to see if they might be due to previously unrecognized blends toward one or both of the atlas stars. As will be discussed in Section 6, such differences are larger than those that might be related to the differences in the interstellar component structure observed in higher resolution spectra of those two stars, and they can be evaluated by examining the spectra of the other twenty-three program stars of this work. The most repeatable solution, regarding whether or not to subdivide the absorption profile and how, was deduced and served as the proposed update (see examples below). After that, measurements of the subject feature were made using the semi-automated program *arcexam* (see Section 4.3) following both the original Atlas Catalog and the proposed tentative update for the APO Catalog of DIBs. Finally, the proposed update was accepted if it provided more consistent measurements and results than the comparison from the two results of the original atlas stars (where the measurements were originally made by hand). We emphasize again that by "being consistent", we refer to a set of measuring techniques (continuum placement and boundary setting) that can be applied to most of the selected spectra. The separation point for possible blended DIBs is not fixed to a particular wavelength, but to a specific feature within the spectrum, such as the inflection point consistent with the independently set continuum level in the region.

Among the updates suggested for the Atlas Catalog of Hobbs et al. (2008 and 2009), we found the most common cases to be DIBs with similar wavelengths but quite different FWHM in the atlas stars. Keeping in mind that we do not know the intrinsic profile of any given DIB or the extent to which that profile can vary in different environments, the solution is usually to split the broader DIB into multiple narrower features, each of which has similar profiles in the multiple program spectra being examined for in this work. This is similar to the approach used by Porceddu et al. (1991), who concluded that DIB $\lambda$6203.6 was a combination of two separate DIBs. One such example in this work is the DIB at 4501.8Å. Its FWHMs are reported to be 3.01 and 2.05Å in HD 183143 and HD 204827 respectively in the Atlas Catalog (column H, index lines 6 and 7 of Table 2). This DIB has a broader profile in HD 183143 than in HD 204827, because of the structure centered at ~ 4504.5Å beside the dominating absorption (top and 2nd spectrum of the left panel of Figure 1). A similar structure is in fact seen in the spectrum of HD 204827, but could be due to the stellar contamination (Hobbs et al. 2008). However, by plotting the same wavelength region in more spectra, we found the same structure in other sight lines whose comparison stars suggest no stellar contamination. An inflection point can be found around 4503.4Å and can be very close to the continuum level in some cases (e.g. HD 43384, 3rd spectrum in the same panel). Our evaluation of this region thus indicated that splitting this broad feature into two separate DIBs at $\lambda\lambda$4501.5 and 4504.5Å would be more appropriate. The (average) FWHMs of the two DIBs are 2.53Å and 0.83Å, respectively (columns A and E, index numbers 6 and 7 of Table 2). This correction could not have been made with evaluation of a single spectrum for this DIB, even if it has a very high S/N ratio.

Another example involves the two DIBs between 7482 and 7485Å (Figure 1 right panel). Two narrow DIBs, centered at $\lambda\lambda$7482.9 and 7484.2Å and with FWHM = 0.61 and 0.55Å (respectively), were reported for HD204827. On the other hand, a single, much broader DIB (FWHM = 2.23Å) was identified at 7483.3Å toward HD 183143. In the Atlas Catalog, the DIBs $\lambda\lambda$7482.9 (from HD 204827) and 7483.3Å (from HD 183143) were considered as a common DIB, despite their very different widths (columns D and G, index lines 521 and 522 of Table 2). By checking the expanded spectral sample of this work, we found two narrow features with similar central wavelengths and FWHMs as observed towards HD 204827 in most of the program sight lines. An inflection point is usually found around 7483.5Å and is also seen in the spectrum of HD 183143, but did not reach the local continuum level. In Hobbs et al. (2008 and 2009), DIBs that appeared to have structure were considered as one DIB if the high point in the feature did not reach the continuum. This convention led to the identification of one broader DIB instead of two narrower ones in HD 183143 for the Atlas Catalog. We concluded that two narrow DIBs at $\lambda\lambda$7482.9 and 7484.2Å (as identified in HD 204827) are present in this region, and they are found in both atlas stars.

Sometimes, it can be very hard to properly separate an absorption complex without referring to spectra of multiple sight lines. The weaker of two blended DIBs can be easily taken as a sub-structure or wing of the stronger DIB, especially in highly reddened sight lines, and vice versa. The genuine separation point can also be blurred due to severe blending, or be hidden in the noise fluctuation or continuum placement uncertainties. For weak DIBs, it is almost impossible to distinguish a real inflection point between two features from noise based on a single spectrum.

On the other hand, the use of multiple program spectra allows us to check the repeatability of profiles of the DIB. If an inflection point is always found at a relatively fixed wavelength in multiple program sight lines, we can then tell that it is real rather than random. This is crucial for resolving the possible structures within a "single DIB" and finding the inflection point to divide them. But we may not be able to confirm if the structures are blends of DIBs from different carriers, or from multiple transitions of a single carrier (see Figure 1 for examples; note that these structures identified in our spectra are different from the sub-structures observed in, e.g. DIB $\lambda$5797.2 in ultra-high resolution spectra; Sarre et al. 1995). Such problem may be settled by the use of spectra of higher resolution and testing the correlations between these structures, or ultimately after specific carriers are assigned to a particular absorption feature. Despite the adjustments made in this work, observations at higher resolution or the use of new techniques may lead to slightly different definitions of these DIBs. For example, Bernstein et al. (2018) proposed that DIB $\lambda$6613.6 contains a doublet and a red-tail component, where the doublet shares the same carrier with DIB $\lambda$6196.0, and the red-tail is due to another carrier. This result is based on spectra of very high resolving power (R ~ 220,000) and these structures cannot be resolved in our data.

4.2 Searching for New DIBs

We searched for new DIBs in the selected sight lines following the classical routine (see e.g. Galazutdinov et al. 2000; Weselak et al. 2000; Tuairisg et al. 2000; Hobbs et al. 2008 and 2009). Each program spectrum is examined by eye in a 50Å-wide window using the plotting function of our online database[4], along with its comparison star which has similar spectral type, small projected rotational velocity and minimal reddening (Table 1). The spectrum of the comparison star is used as reference for the stellar lines which could present in the program star spectrum. The spectrum of a telluric reference star, 10 Lac, from which telluric lines have not been removed (labeled as 10 Lac-tel in our database) is also displayed to provide guidance for locations with possible telluric residuals. Known DIBs are marked with vertical bars at the central wavelengths, and the spectra of HD 183143 and HD 204827 are used for DIB profile reference. If observations on multiple nights were available, these spectra (of each separate night) are displayed along with the sum of all available spectra of the program star.

The primary requirement of the DIB candidates is that the feature should have relatively fixed central wavelength, FWHM, and profiles in all nightly spectra as well as the co-added final spectrum. For each of the sight lines thus visually examined, the estimated central wavelength and FWHM of all new DIB candidates are recorded in a temporary list.

The possible new DIBs were then crossed matched within all the program sight lines. A semi-final list of new DIB candidates was then compiled for the candidates detected in multiple program sight lines with similar central wavelengths and profiles. Additional new DIB candidates came from the "possible DIBs" in HD 183143 listed in Table 3 of Hobbs et al.

---

[4] see http://dib.uchicago.edu/

(2009). There has been no report that HD 183143 is a binary system (Chentsov, 2004; Hobbs et al. 2009), and multiple searches for signs of a binary system were carried out by D.G.Y. in 2014 without success. These "possible DIBs" had about the same width as the rotational widths of stellar lines of HD 183143 and could not be unambiguously confirmed using spectra of HD 183143 alone.

Measurements of the possible new DIBs in this semi-final list were then made with *arcexam* (see the next section) for all program sight lines. For the final confirmation, we require repeated detections of the new DIB candidates in at least 5 sight lines (20% threshold), and that their strengths (represented by EW) be positively related to the reddening. Details of these tests are given in Section 5.

4.3 Measuring with *arcexam*

All measurements in this work were made using *arcexam*, a semi-automated routine written by D. E. W. The display of the spectra in arcexam is similar to that of the plotting function of our online database. For each of the measurements, a small section of the program spectrum around the DIB was displayed along with spectral segments from the atlas stars (HD 183143 and HD 204827) and 10 Lac-tel. These spectra were used to provide reference regarding the expected DIB profiles and to identify possible residuals from imperfect telluric line corrections. The spectrum of the comparison star (Table 1) is displayed to identify possible blends from stellar absorptions. A DIB would be deemed unmeasurable and rejected from the measurement if a strong stellar contamination was discerned. Otherwise, if the possible stellar contamination is minor compared to the DIB, we carried out the measurement and marked the measurement with a "s" flag. Such measurements were excluded from the analysis described in the following sections.

An initial automatic fit to the local continuum is provided at the beginning of the measuring process of each DIB. The user can make manual adjustments if the automatic fit is not acceptable (e.g. when points in stellar lines or telluric residuals were used for continuum placement and the outcome continuum contains too much curvature). Then the extent of the absorption is determined based on the continuum, and again the user can choose to accept or change end points of the DIB absorption vis-à-vis the continuum (Welty et al. 2019, in preparation).

We assume no specific profile for any of the DIBs being measured, and use the direct integration under the continuum within the end points as the EW of the DIB. The uncertainty in the EW is then calculated based on the width of the DIB and the fluctuations in the local continuum region on both sides of the DIB integration window (Jenkins et al. 1973; Savage & Sembach 1991), and a 3-σ criterion is adopted for DIB detection. If a DIB is detected, *arcexam* calculates its FWHM[5] and central wavelength, as well as three additional factors labeled as xm2, xm3, and xm4. The three xm-factors are apparent optical depth-weighted moments of the profile, and are related to the width, skewness, and kurtosis[6] of the absorption profile respectively. The final measurements for each star were incorporated into a visual presentation of the spectra of all the stars, in the Appendix (Figure A1).

For the more general DIB survey project, we have now measured and examined about 250 selected DIBs toward about 180 stars from our current database of about 450 stellar spectra (Thorburn et al. 2003; Friedman et al. 2011; Welty et al, in preparation). In this process, we visually examine wavelength regions from five to ten times the widths of the DIBs measured for the presence of contaminating stellar lines, contaminating telluric lines, and any other DIB-like features. The measurements are evaluated independently and quantitatively, by inter-comparison of the behavior of each DIB in large numbers of stars and by inter-comparison of repeated measurements by different measurers (York et al., in preparation). For the purpose of this paper, the error of EW can be regarded as at a 5 sigma level of 6mÅ for most lines narrower than 1Å, increasing linearly with line width up to 6Å, and are uncertain in different ways for broader lines, due to our use of echelle spectra (see also Section 5.2). The smallest 1-σ errors for the narrowest lines are typically several tenths of a mÅ. More details of the measuring techniques will be included in upcoming papers (Welty et al., York et al., in preparation).

4.4 Detection Limits of the Spectra

The efficiency of detection in this work is constrained by different effects and mechanisms. At short wavelength, the poor detection limits are primarily due to the presence of many stellar lines, which is clearest at $\lambda <$ ~4,500Å for early B stars and at $\lambda <$ ~5,000Å for A stars, and becomes progressively more severe for cooler stars. The threshold for DIB detection is also raised by increasing extinction and by decreasing instrumental sensitivity towards shorter wavelengths.

On the other hand, towards the red part of the spectrum ($\lambda >$ ~7,000Å), the telluric bands become the major obstacle to DIB detection. The telluric lines are carefully corrected, and we visually compare the residuals as they appear in our co-added and telluric corrected spectra of the program star, with the uncorrected telluric reference spectrum (10 Lac-tel, see Section 4.3). Their residuals from slight mis-corrections can still contaminate the spectrum in some cases by affecting the placement of the continuum, which in turn contributes significantly to the uncertainties of DIB EWs. At even longer wavelengths ($\lambda >$ ~8,000Å), usually beyond the wavelength limits of the APO Catalog of DIBs region, the flat-fielding (and thus the continuum determination) are compromised by the appearance of interference fringes from the CCD detector.

---

[5] FWHM is calculated as the distance between the two points at half of the maximal depth of the profile. Please refer to Section 5.2 of this work and discussions in Hobbs et al. (2008 & 2009).

[6] Central wavelength $\lambda_0 = \int_{\lambda_2}^{\lambda_1} \lambda * \tau(\lambda) \, d\lambda / T$; xm2= sqrt$[\int_{\lambda_2}^{\lambda_1} (\lambda - \lambda_0)^2 * \tau(\lambda) \, d\lambda / T]$; xm3 = $\int_{\lambda_2}^{\lambda_1} (\lambda - \lambda_0)^3 * \tau(\lambda) \, d\lambda / (T * \sigma^3)$; xm4 = $\int_{\lambda_2}^{\lambda_1} (\lambda - \lambda_0)^4 * \tau(\lambda) \, d\lambda / (T * \sigma^4)$. Here $\lambda_1$ and $\lambda_2$ refer to the integral limits of the DIB profile, $\tau(\lambda)$ is the apparent optical depth, and $T = \int_{\lambda_2}^{\lambda_1} \tau(\lambda) \, d\lambda$ is the sum of the optical depth.

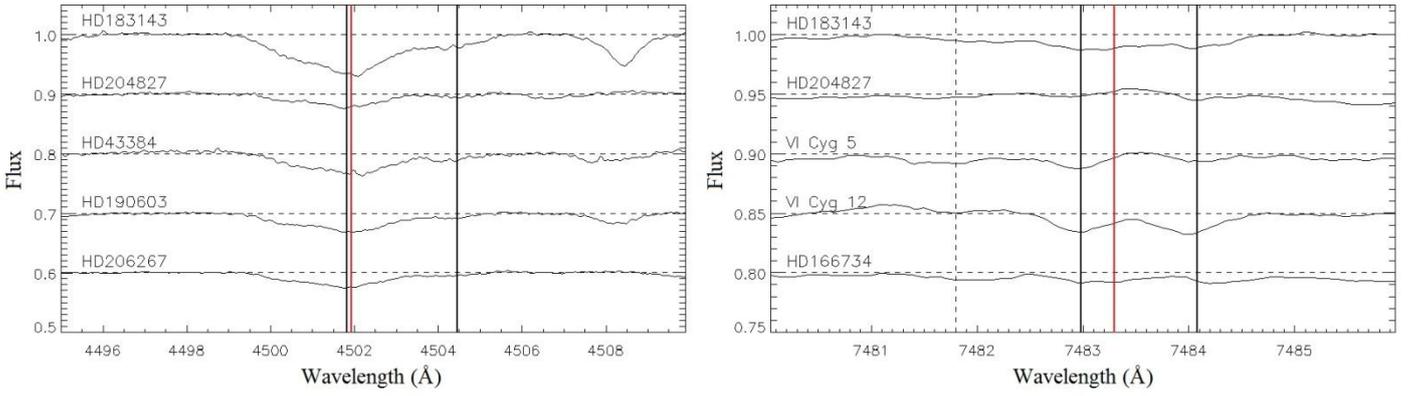

Figure 1. Examples of DIBs for which small changes in wavelengths and widths were made between the Atlas Catalog and the new APO Catalog of DIBs. Portions of the spectra of five stars from Table 1, including HD 183143 and HD 204827, are given in each panel. Left panel: the red vertical line denotes the $\lambda$4501.8 DIB listed in the Atlas Catalog (Table 2, columns E and H, index 6 and 7) whose FWHM is quite different in HD 183143 and HD 204827. Inspection of the twenty-five spectra revealed that splitting this DIB into two DIBs at $\lambda\lambda$4501.5 and 4504.5Å, as marked by the black lines and as explained in Section 4.1, would be more appropriate. Right panel: in the Atlas Catalog, a single DIB was identified in HD 183143 at 7483.3Å (top spectrum of this panel and the red vertical line), and two narrower DIBs were reported at $\lambda\lambda$7482.9 and 7484.2Å in HD 204827 (2nd spectrum and the black vertical lines). From the examination of additional stars, we found the "broader DIB" in HD 183143 to be the blend of the two narrower features. The DIBs originally reported in HD 204827 are retained in our new APO Catalog of DIBs as the best solutions for the 14 and 10 detections, respectively, for these two features in the twenty-five program stars. The dashed vertical lines are DIBs that appear within the plotting windows.

# 5 Results

## 5.1 Detection Percentages of Individual DIBs and Results

Aiming for a self-consistent catalog, we investigate the repeatability of the central wavelength and width of a given DIB in all sight lines as a measurement of the reality of a given DIB. Note that we have certain acceptance for variations in the wavelength width, and end points for a given DIB, rather than forcing these parameters to a very specific value in different sight lines. This is to account for possible physical effects within a given region caused by local conditions (e.g. Dahlstrom et al. 2013). The detection percentage of an individual DIB is defined as the number of detections of this DIB in our program sight lines, divided by twenty-five, which is the total number of program sight lines selected in this work. The histogram of the detection percentages of DIBs in the Atlas Catalog (the aggregate of the results from Hobbs et al. 2008 and 2009) is presented as light gray bars in Figure 2. We have excluded DIBs that do not agree in wavelength and/or width in the two atlas stars, and the remaining 504 DIBs were used in this analysis. The histogram for DIBs in the new APO Catalog of DIBs is presented as diagonal bars in the same figure.

The detection percentages for the 504 DIBs from the Atlas Catalog being tested range between 4 and 100%, with the median value of 52% (note the detection in each of the twenty-five program sight lines contributes a 4% of the detection percentage). However, the variation of the number of points among the bins is rather smooth and no obvious edge or cutoff can be found in the histogram (gray bars in Figure 2). Many of the DIBs with very low detection rates are seen primarily toward the most reddened stars. We set a minimum number of detections for a DIB to be considered real for the APO Catalog of DIBs at 20%, which would omit ~5% of the DIBs being studied (see the accumulated curve in Figure 2 with the inverted ordinate at the right side of the figure). The requirement of a minimum number of detections across the twenty-five stars in the APO Catalog of DIBs is meant to strengthen our confidence that extraneous detections are not included as official DIBs. Although this criterion is somewhat arbitrary, DIBs with only a handful of detections would result in even fewer clean measurements (measurements without contaminations from any sources).

As an example of the type of errors that necessitate a cutoff for the number of stars in which a DIB is detected, consider that "perfect" spectral comparison stars with low reddening and low *vsini* are hard to find. Abundance differences between stars may produce cases in which a stellar line appears in the program star but not in its comparison star. In a classical routine the most straightforward conclusion may be that this feature is not, after all, stellar. This choice would then produce a false DIB. Such an error would be recognized when this "false

DIB" is not detected in other program stars of different spectral types in this study. Thus, confirmation of each real DIB is essential.

We examined the DIBs with insufficient detections in our more extended data base containing measurements of 232 spectra by the date of this work. The number of detections of these DIBs range between 5 and 34, corresponding to the detection percentages between 2.2% and 14.7%. These detection percentages are similar to what are reported in this work. So it is possible that some of the features with relatively low detection percentages are genuine but rare DIBs, while the features with very limited detections in the more extended data sample are most likely not real. But for the reason of being consistent within this work, we still consider these DIBs "unconfirmed".

The correlation between the EW of each DIB and the $E_{B-V}$ of the sight line is also tested. For each of the DIB being considered, we use all clean measurements made in the twenty-five program spectra and normalize these measurements by the largest EW we obtained for this DIB. These normalized EWs are fitted to a linear regression with the reddening of the sight lines by least square method assuming both the uncertainties in the EWs of the DIBs as well as in the $E_{B-V}$ values. A general growing trend is expected in the comparison, as specified by the interstellar origin of the DIBs. The DIBs being examined showing negative or close-to-zero slopes are considered " unconfirmed ". The intercepts of the best-fit lines for DIBs with positive slopes are typically smaller than 0.3, an after-the-fact criteria; there is no requirement regarding how good the correlation is (e.g., as given by the linear correlation coefficient $r$). In total, twenty DIBs have unsatisfying correlations with $E_{B-V}$, many of which have very limited points (typically < 5 points). We found no robust negative slopes for any DIB when compared to the reddening (Baron et al. 2015).

The 559 DIBs that conform to the two criterion (of a minimum detection percentage in the twenty-five stars, and show a generally growing trend with $E_{B-V}$) are listed in Table 2, which served as our "*Apache Point Observatory Catalog of Optical Diffuse Interstellar Bands*" (APO Catalog of DIBs). As we are using spectra from multiple sight lines, average values over the detections from the twenty-five stars are used as the final results for the central wavelengths (Column B), FWHMs (Column F), and the xm2 factors (Column I; see Section 4.3). Their corresponding uncertainties are represented by the standard deviation (SD) values, as listed in Column C, D, G, and J, where the SD of wavelength is given in the unit of Å as well as km s$^{-1}$. We also report the average normalized EWs (by $E_{B-V}$) of each DIB as their typical strength in Column K. To ensure the quality of the values reported in the APO Catalog of DIBs, we restricted ourselves to only the clean measurements free from any apparent contaminations by stellar lines or telluric lines. The central wavelengths and FWHMs reported in Hobbs et al. (2008 and 2009; essentially the Atlas Catalog) are listed in Column E and H, respectively, for comparison. The numbers of detections, limits (measured but not detected), and rejections (not measured for various reasons) are listed in Column L. These numbers lead to the two detection percentages reported in Column M, where the "all detection percentage" is the number of detections divided by 25 (the total number of program sight lines), and the "measured detection percentage" is the number of detections divided by the sum of detections and limits (the measurable portions of the twenty-five program spectra). The main entries of the new APO Catalog of DIBs are columns B, F, and I of Table 2.

However, both the necessary number of confirming detections for a given DIB, and the "apparent" growing trend for its EW with reddening, are somewhat arbitrary. We therefore list in Table 3 of this work the wavelengths and FWHMs of the features that failed either of the two tests, on the detection percentage and $E_{B-V}$ correlation. These values are taken from the Atlas Catalog, in case they are later found to be just very rare DIBs. However, it is interesting to note that, none of these features are considered "common DIBs" in the Atlas Catalog (see columns B and C of Table 3 which indicate the detections of all of these DIBs in only one of the atlas stars). Table 3 also lists the particular criterion a given DIB failed. The broad DIBs listed as possible in Sonnentrucker et al. (2018, hereafter S18) are also given in Table 3 since further confirmation is needed for these DIBs.

The efficiency of the detection of DIBs is subject to various observation factors (Section 4.4). In Figure 3, we summarize the number of DIBs detected in 200Å-wide wavelength windows between 4,000Å and 9,000Å. The wavelength regions where telluric regions are corrected are noted as diagonal areas in the figure. It is not surprising to find that the most sensitive wavelength region for DIB detections is between approximately 5,500 and 7,000Å. The sharp decreases in the number of DIBs at $\lambda > 7,500$Å and $\lambda < 5,000$Å are partly due to the presence of telluric bands and more stellar lines, and our list could be incomplete in these regions, especially for weak DIBs. But with the high S/N ratios of our spectra, the dearth of DIBs in those regions could also be real. Observation towards fast-rotating (very large *vsini* value) early type star may help to ease the obstacles from stellar lines especially for $\lambda <$ ~5,000Å, and additional DIBs may still be found with the use of spectra with even higher S/N ratios, observations from space that avoid strong telluric lines (like the A-band), studies such as this one with larger spectral sample, or confirmation of some of the possible candidates listed in Table 3 of this work.

## 5.2 Broad DIBs

The identification of broad DIBs has always been a difficult task due to the placement of continuum and unresolved blends with stellar lines and other DIBs. For a given EW, the DIB profile becomes shallower with increasing width. When the central depth of the profile is too small, the profile is very likely to be hidden in the fluctuations due to noise and difficult to distinguish from the continuum. The use of medium-high resolution echelle spectra also contributes to difficulties for the broad DIBs. Spectra extracted from each order suffer a curvature owing to the blaze function of the spectrograph. This leads to the uncertainties in the global continuum, especially when a DIB straddles the joining point between adjacent orders. A threshold of ~6Å was adopted by Hobbs et al. (2008 and 2009) as the limit for the FWHM for which the continuum can be accurately determined in the ARCES spectra. The spectra used in this paper were taken from the same data collection.

While more narrow DIBs could in principle be discerned in the echelle spectra as used in this work, the detections of broad DIBs are uncertain given the difficulties listed above. Thus, we refer to the broad DIB catalog of Sonnentrucker et al. (2018, S18). This catalog is based on a specially designed homogeneous survey of broad DIBs using R ~ 500 spectra, and contains nine confirmed cases and eleven possible candidates with FWHM > 6Å.

While examining the twenty-five program stars of this work, we detected six of the nine confirmed broad DIBs from S18. All of these nine confirmed broad DIBs are included in Table 2, regardless of their numbers of detections in our data sample. We also detected two broad DIBs listed as "possible" in S18 and these are included in Table 2. Thus, eleven DIBs in Table 2 of this paper come uniquely from S18 (two possible broad DIBs plus all the nine confirmed broad DIBs in S18), eight of which are independent detections in this work. The rest of the possible broad DIBs in S18 are summarized in Table 3 of the current paper as possible DIB candidates for completeness.

However, despite the identification of the broad DIBs, their true continua are hard to determine in our spectra. The measured EWs as well as other parameters (central wavelengths, FWHM, and the xm2 factors) for the broad DIBs are thus subject to larger uncertainties as illustrated in columns C, D ,and G, and J of Table 2. Such systematic errors have already been suggested by Hobbs et al. (2008 and 2009). With more stars in the data sample, we will show in Section 6 that the uncertainty of the DIB central wavelength does indeed increase with the width of its profile.

## 5.3 Comparison with Previous Works

5.3.1 Atlas Catalog by Hobbs et al.

The APO Catalog of DIBs largely confirms the detections of Hobbs et al. (2008 and 2009). About 480 DIBs from the Atlas Catalog are confirmed without significant changes in central wavelengths and FWHMs, corresponding to ~85% of the total DIBs reported in Table 2. Another 60 DIBs in Table 2 are refined versions of 27 (sets) of features in the Atlas Catalog. They are from wavelength regions that have already been identified as DIB absorptions in Hobbs et al. (2008 & 2009), but we adopted different boundaries of integration during the process described above. This similarity provides empirical evidence to support the view that the populations in the DIBs toward the two atlas stars includes nearly the full range of DIBs to be found in the solar neighborhood. Only 22 truly new DIBs (compared to the Atlas Catalog) have been identified in this work, and 8 of them were among the possible DIBs in HD 183143 (Hobbs et al. 2009; see Section 4.2 of this work). This corresponds to a very small fraction of the total DIBs in the APO Catalog of DIBs, and reflects both our adopted detection limits and our requirement that DIBs should have sufficient detections ($\geq 5$) in our spectral sample. Such requirements help ensure the self-consistency of the catalog, but might also block the confirmation of some rare DIBs found only under certain ISM conditions (that may not be adequately covered in our sample).

We included stars with widely varying physical environments in this work, expecting detections of many new DIBs arise from the products of photon-reactions in sight lines containing intense ISRF (e.g. HD 37061 and HD 37903), or from the precursors of DIB carriers in heavily shielded sight lines (e.g. HD24534). Such attempt is unsuccessful since only a handful of new DIBs are identified in this work. It is possible that these "new molecules" do not have transitions in the range of our optical DIB catalog, or are not abundant enough to be detected. Alternatively, their transitions may have already been noted as known DIBs in the Atlas Catalog.

5.3.2 Other Works

There have been a number of deep surveys aimed at establishing a complete catalog of DIBs. The size of spectral sample, wavelength coverage, and the number and kind of DIBs reported vary for different projects, but all these efforts relied on high quality spectra from carefully selected sight lines (e.g. Jenniskens et al. 1994; Galazutdinov et al. 2000; Weselak et al. 2000; see Sonnentrucker 2014 for a brief summary).

We first compare the APO Catalog of DIBs to the result of Tuairisg et al. (2000), which is based on the observations towards three sight lines, BD+63 1964, BD+40 4220 (Cyg OB2 5), and HD 183143 (the latter two sight lines are also used in this work). The results are in good agreement. For the 226 DIBs confirmed between 3906 and 6812Å in Tuairisg et al. (2000), we found 197 to be listed in Table 2 of this work. Among the 29 DIBs that are not included in our catalog, 11 are broad DIBs with FWHM > 6 Å, and some of them are included in Table 3 as possible DIBs.

A more recent DIB atlas containing 336 DIBs between 3,500 and 10,000Å presented by Bondar (2012) is based on the observations of ten reddened O and early-B type stars, with stellar models used to identify the stellar lines. The spectra were from multiple instruments and of higher resolving power (~100,000) but somewhat lower S/N ratio (typically 500 around 6,000Å) than this work. In total, 301 DIBs in Table2 of this work have their counterparts in the Bondar atlas (note that they may not be one-to-one correspondence since the separation of some DIBs were handled differently in the two works). Many of the 258 DIBs reported in this work that are not included in the Bondar atlas are located in wavelength regions affected by telluric absorption bands, as telluric reference observations apparently were not always available to aid in identifying the DIBs. Another 41 DIBs in the Bondar atlas were not included in our APO Catalog of DIBs. We found most of these DIBs are "possible", and have a relatively small number of detections (as reflected by the number of sight lines used to obtain the average wavelength and FWHM; Table A2 of Bondar 2012). Some of them are included in our Table 3 as possible DIB candidates.

5.4 The DIB Populations in the Atlas Sight Lines

As noted in Section 1, Hobbs et al. (2008, 2009) found a rather different mix of DIBs in the two atlas stars: 165 DIBs uniquely found in HD 183143; 131 DIBs uniquely found in HD 204827; and 249 DIBs common to both. One of the prime motivations for this paper was to see if additional diversity in the DIB population would be revealed when more stars were investigated.

However, the apparent diversity of DIB population in the atlas sight lines is less strong then it first appeared in the Atlas Catalog. Among the 559 DIBs, we found 416 detected in both of the atlas sight lines at 3-$\sigma$ level. There are 56 DIBs uniquely detected towards HD 183143, 75 towards HD 204827, and 12 not detected in either of the two sight lines. The sets of the DIBs detected in the two atlas sight lines are very similar, despite the very noticeable differences in ISM conditions found along them (Hobbs et al. 2008, 2009, and references within). The increase in the number of the common DIBs of the two atlas stars might be a combination of the following factors,

1) Some DIBs previously considered to be unique to one atlas stars appear to be weakly present in the other as well, and some of the possible DIBs found in HD 183143 (Table 3 of Hobbs et al. 2009) were confirmed in HD 204827.

2) The adjustments unified some unique DIBs for the two atlas stars.

3) Some common DIBs were divided into multiple DIBs during the adjustments.

4) The contribution from the new DIBs found in this work.

So the diversity of DIB populations is related to the DIB catalog being used, and may be a matter of relative strength rather than of presence or absence. The large databases now being constructed (e.g. the EDIBLES survey, Cox et al. 2017), and correlations between the DIBs and independently determined physical conditions will shed light on these matters in the future years.

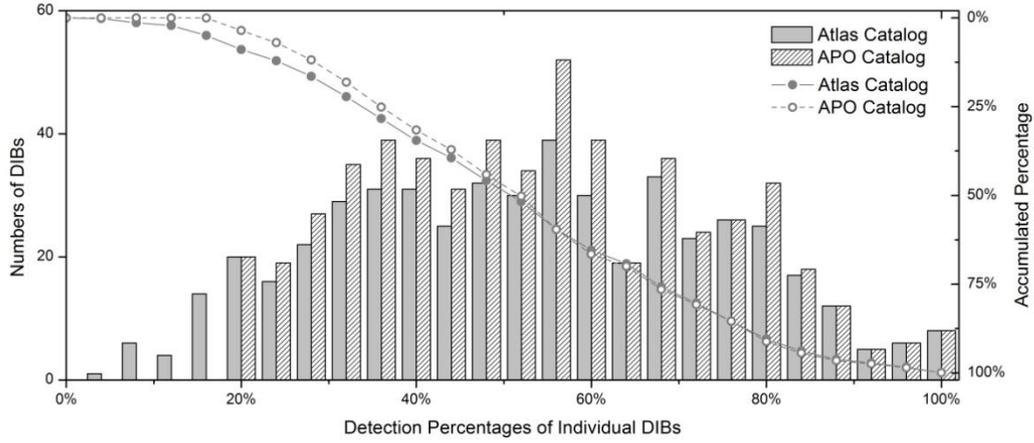

Figure 2. Histogram of detection percentage of each DIB. This percentage is defined as the number of DIB detections divided by 25, the number of program sight lines. The solid vertical gray bars are for the previous Atlas Catalog (from Hobbs et al. 2008 and 2009), and the vertical bars filled with diagonal lines are for the new APO Catalog of DIBs (Table 2 of this work), respectively. We also report their accumulated percentage curves labeled by filled and empty circles. Note that the two samples are largely overlapped, resulting in the very similar patterns seen in the plot. The sharp edge for the diagonal bars at 20% is due to the minimum detection percentage adopted for the APO Catalog of DIBs.

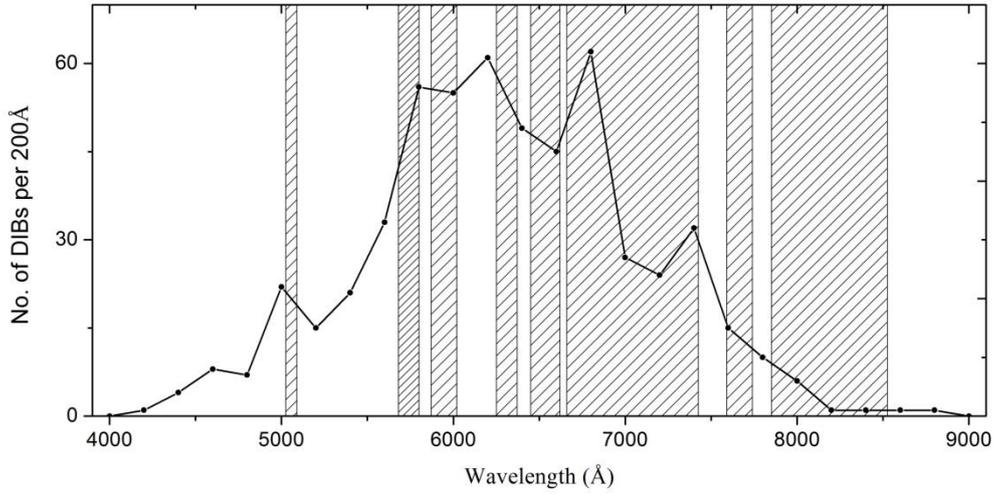

Figure 3. Number of DIBs detected in 200Å-wide wavelength intervals between 4,000Å and 9,000Å. The shaded areas mark regions where telluric lines have been corrected. The highest densities of DIBs are between ~ 5,500 and ~ 7,000Å due to various observational biases, e.g. the contaminations from stellar lines at $\lambda < 4,500$Å and from telluric residuals at $\lambda > 7,000$Å, as summarized in Section 5.4.

Table 2. 559 Diffuse Interstellar Bands Identified in this Work[a]

| (A) | (B) | (C) | (D) | (E) | (F) | (G) | (H) | (I) | (J) | (K) | (L) | (M) | (N) |
|---|---|---|---|---|---|---|---|---|---|---|---|---|---|
| No. | Avg. $\lambda_c$ | SD of $\lambda_c$ | SD of $\lambda_c$ | $\lambda_c$ in the Atlas Catalog[b] | Avg. FWHM | SD of FWHM | FWHM in the Atlas Catalog[b] | Avg. xm2[c] | SD of xm2 | Avg. EW/$E_{B-V}$[d] | No. of Det/Lim[e]/Rej[f] | Det. Pct. All[g]/Mea.[h] | Comments |
| | (Å) | (Å) | (km s$^{-1}$) | (Å) | (km s$^{-1}$) | (km s$^{-1}$) | (km s$^{-1}$) | (km s$^{-1}$) | (km s$^{-1}$) | (mÅ mag$^{-1}$) | | (%) | |
| 1 | 4259.00 | 0.16 | 11.44 | */4259.01 | 83.74 | 30.53 | */74.0 | 31.31 | 11.53 | 9.72 | 14/3/8 | 56.0%/82.4% | |
| 2 | 4363.83 | 0.04 | 2.52 | */4363.86 | 41.19 | 11.06 | */31.6 | 13.50 | 2.97 | 7.64 | 12/12/1 | 48.0%/50.0% | |
| 3 | 4371.60 | 0.19 | 12.73 | 4371.73/* | 62.31 | 16.08 | 70.7/* | 24.84 | 3.80 | 6.99 | 7/6/12 | 28.0%/53.8% | |
| 4 | 4429.33 | 1.34 | 90.63 | 4428.83/4428.19 | 1634.62 | 357.28 | 1528.0/1523.9 | 631.92 | 89.32 | 2005.95 | 20/3/2 | 80.0%/87.0% | S18[i] Confirmed |
| 5 | 4494.53 | 0.38 | 25.62 | 4494.55/* | 142.17 | 52.13 | 139.5/* | 49.58 | 9.94 | 13.89 | 8/11/6 | 32.0%/42.1% | |
| 6 | 4501.51 | 0.16 | 10.49 | 4501.66/4501.79 | 168.92 | 22.99 | 200.6/136.6 | 60.60 | 7.54 | 73.84 | 18/1/6 | 72.0%/94.7% | Split into two DIBs |
| 7 | 4504.45 | 0.18 | 11.99 | | 54.95 | 24.77 | | 23.51 | 5.86 | 11.64 | 12/11/2 | 48.0%/52.2% | |
| 8 | 4659.86 | 0.05 | 3.40 | */4659.82 | 35.19 | 4.96 | */29.0 | 13.16 | 1.85 | 4.70 | 10/12/3 | 40.0%/45.5% | |
| 9 | 4668.66 | 0.12 | 7.75 | */4668.65 | 38.82 | 19.94 | */43.1 | 18.85 | 3.71 | 9.91 | 14/6/5 | 56.0%/70.0% | |
| 10 | 4680.24 | 0.07 | 4.27 | */4680.20 | 45.83 | 7.83 | */41.7 | 18.11 | 2.63 | 8.65 | 8/16/1 | 32.0%/33.3% | |
| 11 | 4683.03 | 0.06 | 3.90 | */4683.03 | 29.21 | 3.62 | */27.5 | 11.06 | 1.63 | 9.30 | 18/7/0 | 72.0%/72.0% | |
| 12 | 4688.84 | 0.07 | 4.78 | */4688.89 | 26.87 | 11.57 | */30.7 | 10.88 | 1.63 | 5.22 | 7/17/1 | 28.0%/29.2% | |
| 13 | 4699.29 | 0.16 | 10.03 | 4699.21/* | 116.61 | 33.58 | 86.8/* | 37.24 | 6.89 | 9.96 | 5/11/9 | 20.0%/31.3% | |
| 14 | 4726.98 | 0.16 | 10.36 | 4727.16/4726.83 | 174.89 | 22.34 | 197.4/173.9 | 83.50 | 7.97 | 168.87 | 24/0/1 | 96.0%/100.0% | |
| 15 | 4734.77 | 0.08 | 4.83 | */4734.79 | 26.31 | 8.20 | */26.6 | 10.73 | 1.89 | 7.26 | 17/8/0 | 68.0%/68.0% | |

a: full table available as online supplement;
b: in the format of HD183143/HD204827, values taken from Hobbs et al. (2008 and 2009);
c: optical-depth weighted variance of the DIB profile as reported by *arcexam*. This quantity measures the width of the profile, see Section 4.3 for details and Section 6.2 for discussions;
d: based on measurement specially made for this project, subtle differences are expected with the public database (e.g. Fan et al. 2017)
e: measured but not detected;
f: cannot measure due to, e.g., stellar contamination, residual from telluric correction, or overly complicated continuum;
g: number of detections divided by 25, the total number of program sight lines used;
h: number of detections divided by the sum of detections and limits, that is, the measurable portions of the twenty-five program spectra of this DIB;
i: Sonnentrucker et al. (2018).

Table 3. List of 42 Possible Narrow and Broad Diffuse Interstellar Bands

| (A) No. | (B) $\lambda_c$ 183143/204827[a] | (C) FWHM 183143/204827[a] | (D) Insufficient Detections[b] | (E) Abnormal $E_{B-V}$ Correlation[c] | (F) Comments[f] |
|---|---|---|---|---|---|
| 1 | 4176 [d] | 23.33 [d] | o | | S18 possible, T00 |
| 2 | 4593 [d] | 28 [d] | o | | S18 possible, T00 |
| 3 | 4650.77/* | 1.61/* | o | | |
| 4 | */4879.96 | */1.58 | | o | B12 possible |
| 5 | 4969 [d] | 33.70 [d] | o | | S18 possible |
| 6 | 5039 [d] | 17.87 [d] | o | | S18 possible, T00 |
| 7 | 5130.36/* | 0.88/* | o | | |
| 8 | */5133.14 | */0.94 | | o | |
| 9 | */5137.07 | */0.43 | | o | B12 certain |
| 10 | */5178.10 | */0.48 | o | | |
| 11 | */5229.76 | */0.50 | o | o | |
| 12 | */5395.64 | */1.11 | o | o | |
| 13 | 5413.52/* | 0.50/* | o | | |
| 14 | */5433.50 | */0.45 | | o | |
| 15 | */5566.11 | */1.41 | o | | |
| 16 | 5674.61/* | 0.54/* | o | | |
| 17 | */5914.79 | */0.38 | | o | B12 possible |
| 18 | */6054.50 | */0.98 | o | | B12 possible |
| 19 | */6082.33 | */0.88 | o | | |
| 20 | 6124.43/* | 1.08/* | | o | |
| 21 | 6133.55/* | 0.80/* | o | o | |
| 22 | 6207 [d] | 11.7 [d] | o | | S18 possible |
| 23 | 6281 [d] | 8.47 [d] | o | | S18 possible, T00 |
| 24 | 6311 [d] | 23 [d] | o | | S18 possible, T00 |
| 25 | 6359 [d] | 37.33 [d] | o | | S18 possible, T00 |
| 26 | 6420.80/* | 0.91/* | | o | |
| 27 | 6451 [d] | 25.4 [d] | o | | S18 possible |
| 28 | */6485.71 | */0.59 | o | o | B12 possible |
| 29 | 6534.54/* | 12.69/* | | o | |
| 30 | */6629.60 | /*0.62 | o | | |
| 31 | */6814.20 | */0.56 | o | o | |
| 32 | 6831.21/* | 0.60/* | o | o | |
| 33 | 6951.81/* | 0.86/* | o | | |
| 34 | 7004.51/* | 0.85/* | | o | |
| 35 | 7152.25/* | 2.26/* | o | o | |
| 36 | 7342.73/* | 1.71/* | o | | |
| 37 | 7476.77/* | 1.11/* | o | o | |
| 38 | 7532.73/* | 1.12/* | o | | |
| 39 | 7569.83/* | 5.61/* | o | | |
| 40 | 7950.77/* | 2.02/* | o | o | |
| 41 | 7968.09/* | 1.86/* | o | o | |
| 42 | 8085.94/* | 2.36/* | o | | |
| 43 | 8772.77/* | 2.36/* | o | o | |

a: values gathered from Hobbs et al. (2008 and 2009) unless otherwise specified;

b: with less than five detections in the twenty-five program sight lines of this work;
c: the EW of this feature does not have a general growing trend with $E_{B-V}$;
d: values taken from Sonnentrucker et al. (2018, also noted as S18).
f: S18 - Sonnentrucker et al. (2018); T00 - Tuairisg et al. (2000); B12 - Bondar (2012).

## 6. Discussion

6.1 Detection Percentages

6.1.1 Detection Percentages of Individual DIBs

The detection percentage of an individual DIBs is defined as the number of detections divided by 25 (the number of program stars used in this work). The histogram of this detection percentage for the 559 DIBs in the APO Catalog of DIBs is plotted in Figure 2. The sharp cut at 20% corresponds to the adopted requirement of minimal detection percentage as discussed earlier in Section 5.1; DIBs failing to meet the minimal criteria are set aside as possible candidates in Table 3.

Because most of the DIBs in Table 2 have their counterparts in the Atlas Catalog, the pattern of the histogram of the APO Catalog of DIBs is very similar to that of the reference DIB sample. The median value of the detection percentage is 52% for DIBs in Table 2. This reflects the fact that many DIBs listed in Table 2 are weak features and thus are not detected in many of our sight lines, even though we have focused on sight lines with medium-to-high reddening.

In total, 19 DIBs have detection percentages greater than 90%, corresponding to 23 or more detections in the twenty-five sight lines. These DIBs are, in the order of decreasing detection percentages, DIBs $\lambda\lambda$4963.9, 5779.6, 5780.6, 5797.2, 5849.8, 6203.6, 6284.1, and 6613.7 (with detection percentages of 100%); 4727.0, 6089.9, 6196.0, 6439.5, and 6660.7 (with detection percentages of 96%); 6376.1, 6379.3, 6449.3, and 6993.1 (with detection percentages of 92%). This list covers most of the well-known strong DIBs with a rich history (e.g. Galazutdinov et al. 2004; Cox et al. 2006, 2007; Friedman et al. 2011; Fan et al. 2017). The high detection percentages make them ideal objects for correlation studies.

6.1.2 Detection Percentages of Individual Sight Lines

The detection percentage of an individual sight line is defined as the number of DIBs detected in the sight line divided by 559, the number of DIBs listed in the APO Catalog of DIBs. This rate does not account for cases where a DIB was contaminated by telluric or stellar lines (emission or absorption). The Column "Det. Pct. - All" in Table 1 lists the detection percentages for each of the twenty-five sight lines. In Figure 4, this detection percentage is compared to three quantities which might affect the detectability of the DIBs: the reddening $E_{B-V}$ as an estimate of the total amount of interstellar material along the sight line; the spectral type of the program star; and the molecular fraction of hydrogen $f_{H2}$ representing the overall ISM conditions along the sight line (e.g., Weselak et al. 2004; Vos et al. 2011; Welty 2014).

It is not surprising that the detection percentage generally increases with the reddening

(Figure 4, top panel). For spectra of a given S/N ratio, the weaker DIBs should be more readily detectable in sight lines containing more interstellar material. The steady growth in the detection percentage with increasing reddening can be represented by a linear relationship for the sight lines with $E_{B-V} < \sim 2$ mag. A slope of 33.1% per magnitude would be obtained for this linear correlation from the top panel of Figure 4. Some of the scatter in this relationship can be ascribed to possible secondary dependences on the spectral type of the background star and/or the molecular fraction $f_{H2}$ (see discussion below). For higher reddening, this trend starts to flatten due to the hard limit of 100% in the detection percentage, so that the detection percentage for the sight line of VI Cyg 12 ($E_{B-V} = 3.31$ mag, the most heavily reddened in our data sample) is similar to that for sight lines with $E_{B-V} \sim 2$ mag. Some DIBs were not detected in those most reddened sight lines, however, due to stellar contamination and/or low S/N ratio at the shorter wavelengths (the latter also a consequence of the significant reddening).

A decreasing trend between the detection percentage of DIBs and the spectral type of the program star can be expected, as a result of the increased density of stellar lines in the spectra of cooler stars. However, this decreasing trend can only be loosely identified in our data (Figure 4, middle panel), as the effects associated with the reddening appear to be stronger. To account for the clear trend between detection percentage and the reddening, we examine the residuals of the detection percentages from the best-fit line described above to look for possible secondary trends. The residuals are then plotted against the spectral type of the program star in Figure 5 (top panel), and the decreasing trend becomes more condensed and easier to identify. While the increased blending with stellar lines in cooler stars appears to have a smaller effect on the detection percentages than the reddening, it is one of the key reasons that most DIB studies have been focused on early type stars.

A lambda-shaped behavior is found for the EWs of some well-studied non-$C_2$ DIBs when compared to the $f_{H2}$ value of the sight line (Fan et al. 2017). Their EWs, whether normalized or not by $E_{B-V}$, first increase then decrease with $f_{H2}$, with a peak at $f_{H2} \sim 0.3$. The normalization of EW would reduce the dispersion among the individual sight lines at similar $f_{H2}$, which could be due to differences in distance and reddening. But this lambda-shaped behavior is not observed among the $C_2$ DIBs. These DIBs seem to be more sensitive to dense cloud conditions and sight lines of higher $f_{H2}$, where the $C_2$ molecules are preferentially detected (Thorburn et al. 2003).

This lambda-shaped behavior is also observed in the direct comparison between the detection percentage and the $f_{H2}$ value of the sight line (Figure 4 bottom panel), especially for the decreasing part at larger $f_{H2}$ where more of the data points in our small sample are located. The DIB population is dominated by weak features (Figure 2) whose detectabilities are closely related to their strengths. As only a small percentage of the DIBs are $C_2$ DIBs, the lambda-shaped behavior might characterize the majority of the DIBs. They would be more easily detected when their strength is near the peak value at $f_{H2}$ ~ 0.3, resulting in the lambda-shaped trend observed between the detection percentage and the $f_{H2}$ value of the sight line. Note the above analysis is similar to the direct comparison between $f_{H2}$ and the un-normalized EWs of DIBs, without considering the reddening or the spectral type of the background star. Their effects can be clearly seen in some of the outliers in the plot, e.g. HD 204827 at $f_{H2}$ = 0.67 and detection percentage of 87.8%. The high detection percentage, compared to other sight lines with similar $f_{H2}$ values, could be due to the large reddening and hot background star of this sight line.

To better study the dependence of detection percentage on $f_{H2}$, a modified detection percentage is used, which is the number of detected DIBs divided by the total attempts of measurements for the sight line. For example, 351 DIBs were detected in HD 20041 and 93 DIBs were not measured for various reasons. The total attempts of measurements would thus be 559 - 63 = 496, and the corresponding detection percentage is 351 / 496 = 75.6%. This modification is an attempt to account for the effect from stellar contamination, as the rejected cases are not considered, and the result is given in the Column "Det. Pct - Mea." of Table 1. To address the dependence on $E_{B-V}$, these modified detection percentages are then compared to the reddening of the sight lines, and their residuals from the best-fit line are obtained. These residuals are plotted against $f_{H2}$ in Figure 5 (bottom panel).

The modified detection percentages reduce the cosmic scatter among sight lines with similar $f_{H2}$ values, just as in the comparison between $f_{H2}$ and EWs (Fan et al. 2017), as seen more clear for the declining trend at $f_{H2}$ > 0.3. On the other hand, we only have three sight lines with $f_{H2}$ < 0.3, and the result may be less well determined for the rising part of the lambda-shaped behavior. To conclude, a larger data sample is needed to study how the detection percentages vary with the $f_{H2}$ of a sight line. We would also like to note that sight lines with similar detection percentage may not contain the same kinds or classes of DIBs. There could be hidden patterns regarding the presence of DIBs at different $f_{H2}$ levels, but to explore these patterns is beyond the scope of this work.

6.2 Width of the DIB Profile

Two parameters reported in Table 2 are related to the width of the DIB profile: the FWHM and the xm2 factor, which is the apparent optical depth weighted variance of the DIB profile (Section 4.3). While the FWHM is often used to characterize the widths of features in spectral studies, it often reflects an assumed or fitted line profile (usually Gaussian or Lorenztian). As we made our EW measurements via direct integration, our FWHM were not obtained by fitting the profiles, but were calculated as the wavelength difference of the two points located at half of the measured maximum depth (Hobbs et al. 2008 and 2009). This "full width at half maxima" is based solely on three points within the DIB profile (the two half-maximum points and the deepest point in the middle of the profile), and is subject severely to the noise fluctuations of these points. On the other hand, the xm2 factor is based on all the points between the integral limits, reducing the effects of the noise.

In Figure 6 (top panel), we compare FWHM to the xm2 factor (both converted to the unit of Å) for all the DIBs reported in Table 2. The only obvious outlier is for the DIB at 5779.6Å (5778.0Å in Hobbs et al. 2008 and 2009), which is the measurement over the whole 5780Å region. This DIB feature has a complex profile, consisting of a relatively narrow and deep central absorption (the DIB $\lambda$5780.6Å) plus a shallow but extended feature which appears to be due to multiple blended components (Hobbs et al. 2008, 2009; Dahlstrom et al. 2013; Sonnentrucker et al. 2018; and references therein). While xm2 took both the narrow peak and broad base into consideration, the FWHM focused only on the central absorption of this feature, thus leading to a small FWHM at large xm2 value. Other than that, a linear trend is well established, with an intercept near 0.0 and a slope of 0.38. This slope is close to the expected ratio of 0.42 between the variance (the xm2 factor here) and the FWHM in a Gaussian profile. The difference is likely due to the fact that the DIB profiles are not Gaussian, and to the optical depth weighting used in calculating the xm2 factor.

The main advantage of the xm2 factor should be the improved accuracy, as the use of more points in the spectrum helps to average out (to some degree) the noise fluctuations. We compare the relative uncertainties (the SD error divided by the average value) of the xm2 factor and of the FWHM (Figure 6 bottom panel), and it is clear that the xm2 factor tends to have smaller relative uncertainties. For the measurements made in this work, the xm2 factor may thus be a more accurate descriptor of the width of the DIB profile. The average value of the xm2 factor and its uncertainty for each of the DIBs are listed in Table 2, and we will use this parameter for the width of the profile of each DIB in the following discussion.

6.3 Standard Deviation of Central Wavelengths

For all the program spectra in this work, including the two atlas stars HD 183143 and HD 204827 (Hobbs et al. 2008 and 2009), the wavelength scale is set to the interstellar frame. This is achieved by setting the apparently strongest component (in our ARCES spectra) of the interstellar K I line at 7698.9645Å (Morton 2003) to zero velocity. Slight differences in the central wavelengths (typically several tenths of an Angstrom) are found for the common DIBs toward the two atlas stars (expect for the DIBs where adjustments were made; see Section 3.1). Such differences are also observed among the twenty-five sight lines in our data sample. For the 559 DIBs reported in Table 2, the average and median values for the standard deviations of the central wavelengths are 7.5 and 5.8 km s$^{-1}$ respectively. The histogram is presented in Figure 7 (upper left panel).

The various atomic and molecular species in the ISM often exhibit complex velocity structure when observed at high spectral resolution (e.g., Welty et al. 1994, 1996, 2001; Crane et al. 1995). Spectra of the two atlas stars obtained at resolutions of 175,000 - 200,000 indicate that the K I $\lambda$7698 absorption toward HD 204827 is due to at least eight components over a velocity range of ~ 15 km s$^{-1}$, dominated by two strong components separated by about 3 km s$^{-1}$ (Pan et al. 2004), and that the K I absorption toward HD 183143 is due to at least ten components over a velocity range of ~ 24 km s$^{-1}$, dominated by two groups of components separated by ~ 15 km s$^{-1}$ (McCall et al. 2002). Our lower resolution ARCES spectra exhibit a single, fairly symmetric K I 'component' toward HD 204827, but two 'components' -- of similar depth, separated by 15 km s$^{-1}$ -- toward HD 183143 (Hobbs et al. 2008, 2009). The measured wavelength of a given DIB will depend on its relative amounts in each component in which it exists. Such quantities, in general, cannot be determined at our resolution, but the overall effects may be observed. By comparing the measurements reported in Hobbs et al. (2008 and 2009), York et al. (in preparation) found a systematic offset in the central wavelengths which appears to be related to the radial velocities of the K I components in the two sight lines. Considering the large velocity dispersion of the major components in the sight line of HD 183143, the central wavelengths measured in this sight line alone may not provide very precise references for the other sight line. This is another reason for us to report the average wavelength of each DIB in Table 2.

Other than HD 183143, seven of the program stars, HD 20041, HD 43384, HD 166734, HD 168625, HD 190603, HD 194279, and HD 223385, show the most velocity structure in the profiles of K I and Na I absorptions in high resolution spectra (provided by G. Galazutdinov, the referee of this work). They also have more than one velocity component in our ARCES spectra. Based on the concern of the referee, we redid the calculation of the mean wavelength, FWHM, and the xm2 factor after omitting all these eight stars, but no significant deviations were found in comparing the results of the two samples. As a further check, we compared the central wavelengths and FWHMs of the common DIBs between our APO Catalog of DIBs and the Bondar (2012) atlas, which focused on sight lines lacking evident Doppler splitting in the interstellar K I line in R ~ 100,000 spectra. The average and median differences of the central wavelengths (APO - Bondar) are both 0.01Å and the standard deviation of the difference is 0.11Å. As for the FWHM ratios (APO / Bondar), the average and median values are respectively 1.15 and 1.09, with the standard deviation being 0.47. Thus to use sight lines containing multiple velocity components appears not to have significantly compromised the accuracy of wavelengths. While these components may be reflected in the profiles of some narrow DIBs such as $\lambda$6196.0, their effects, in general, are secondary compared to the uncertainty in the measurements (see discussions below). This may be due to the fact that we are using moderately-high resolution spectra with R ~ 38,000.

For comparison purposes, we measured six absorption lines from known ISM species[7] in the program spectra. The standard deviations of their central wavelengths range from 2.5 to 4.8 km s$^{-1}$, and we take 3.5 km s$^{-1}$ as the typical value (median value, represented by the dashed line in Figure 6, upper left panel), which is about half of the typical DIB value. Thus the Doppler shift from ISM clouds can only explain part of the uncertainties in DIB wavelengths.

In Figure 7 (the top right and the two bottom panels), three parameters related to the DIB profiles were compared to the fluctuations in their central wavelengths: the xm2 parameter representing the width of the profile; the average normalized EW representing the typical strength of the DIB (Column J of Table 2); and an approximate central depth (CD) based on the average normalized EW and the xm2 factor of each DIB. Assuming Gaussian profiles for each DIB, and with the linear relationship between FWHM and the xm2 factor acquired in Section 6.2, we have:

$$EW = \sqrt{2\pi} * CD * FWHM/2.355 = \sqrt{2\pi} * CD * xm2/(2.355 * 0.38)$$
$$CD = 0.36 * EW/xm2$$

Note this assumption is not true for our measurements of DIBs (Section 4.3), and this approximate central depth could be different from the real central depths observed in the spectra.

There is a roughly square-root relationship between the xm2 factor and the standard deviation of the central wavelength (Figure 7, top right panel). The central wavelength used here is the weighted average within the integral limits (see Section 4.3), and is thus a weighted sum of N terms (where N is the number of points, proportional to the profile width). The variance of the central wavelength would also be roughly proportional to the number of points, and the corresponding standard deviation will go roughly as the square root of the profile width. The difficulties in determining the accurate central wavelengths for broad DIBs have been noted in Hobbs et al. (2008 and 2009). This can be extended to the region of relatively narrow DIBs which is clearly demonstrated here with more data from multiple sight lines.

---

[7] The six absorptions are: CH A-X transitions around 4300Å and B-X transitions around 3886Å; CH$^+$ A-X transitions around 3957Å and 4232Å; and K I doublet at 7665Å and 7699Å.

On the other hand, the dependences on the EW and the surrogate central depth are not clear, as no obvious trend can be found in either of the bottom panels of Figure 7. For the normalized EW comparison, there is considerable scatter. All of the extreme outliers (labeled) are relatively broad features (Table 2) with xm2 larger than 2.5Å or FWHM larger than 6Å. This is also the width limit adopted in Hobbs et al. (2008 and 2009) for denoting broad DIBs, and the threshold suggested by Sonnentrucker et al. (2018) beyond which the DIBs are too broad to be studied in echelle spectra. The large fluctuations in their wavelengths are likely due to their width rather than their strength. As for the central depth comparison, we found DIBs with large fluctuations usually have small central depth but the scatter of points in this region is very large. These outliers are for the broadest DIBs with medium strength, rather than strong DIBs with average width (Table 2).

Thus we conclude that the fluctuations in the measured DIB central wavelength among different sight lines are mostly related to measurement uncertainties that depend on the width of the DIB profile; cloud multiplicity can also play a role in some cases. However, we would also like to note that, this conclusion is based on the analysis of the 559 DIBs reported in Table 2. Many of these DIBs are weak features with less well defined continuum than the stronger features. Use of the central wavelengths of DIBs to trace the velocity components in the ISM may become more feasible if the errors obtained for a larger sample of stars are more thoroughly explored. Such a study is underway for several DIBs seen toward several hundred stars (York et al., in preparation).

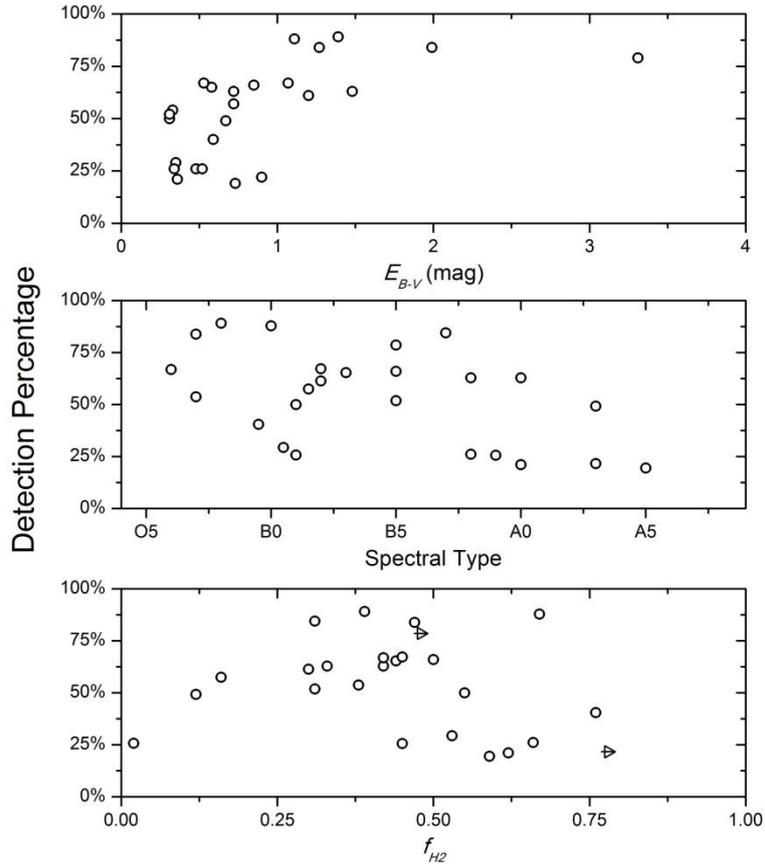

Figure 4. Detection percentages of individual sight lines (number of DIBs detected in the sight line divided by 559) compared to $E_{B-V}$, spectral types of background stars, and $f_{H2}$. An increasing trend can be identified in the top panel, except for the most reddened sight line in our data sample, VI Cyg 12. This is because the detection percentage cannot exceed 100%. There may be a declining trend in the middle panel, and lambda-shaped trend (see Fan et al. 2017) in the bottom panel, especially for the declining trend at $f_{H2} > \sim 0.3$. But these possible trends may be produced by the stronger dependence on the reddening of the sight line. The detection percentage never reaches 100% in our data sample even for the most reddened sight lines due to various reasons such as stellar contamination and low S/N ratio.

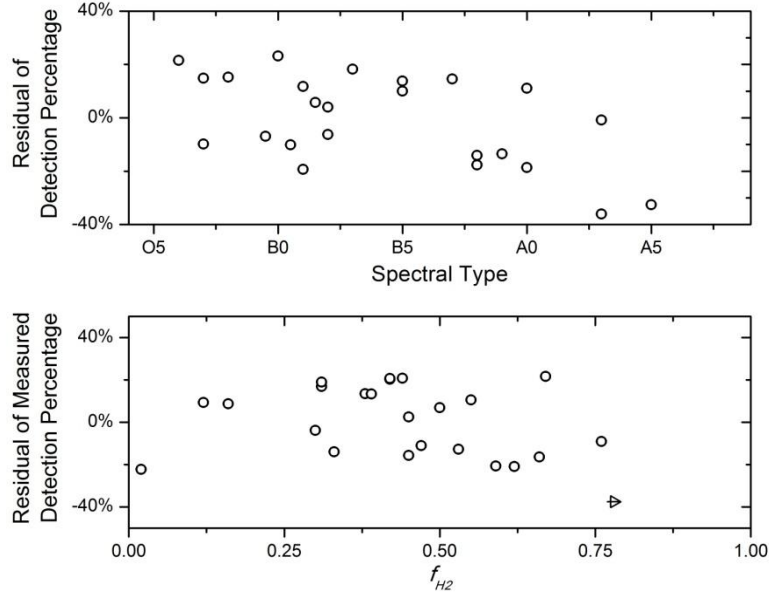

Figure 5. Top panel: the $E_{B-V}$ corrected detection percentage compared to spectral types of the background stars. The $E_{B-V}$ correction was made by converting the detection percentage to unit $E_{B-V}$ following the slope of 33.1% / mag as found in the top panel of figure 4 (See also Section 6.1.2). The declining trend with spectral type is more concentrated after this conversion. Bottom panel: the adjusted detection percentage compared to $f_{H2}$. We calculated the number of DIBs detected divided by the total measuring attempts made in the sight line (559 minus the number of rejections), and then converted it to the unit $E_{B-V}$. The lambda-shaped trend can be identified and the dispersion in the declining trend for $f_{H2} > \sim 0.3$ is reduced after the conversion. As we only have three stars with $f_{H2} < 0.25$, the increasing arm at small $f_{H2}$ is less well determined.

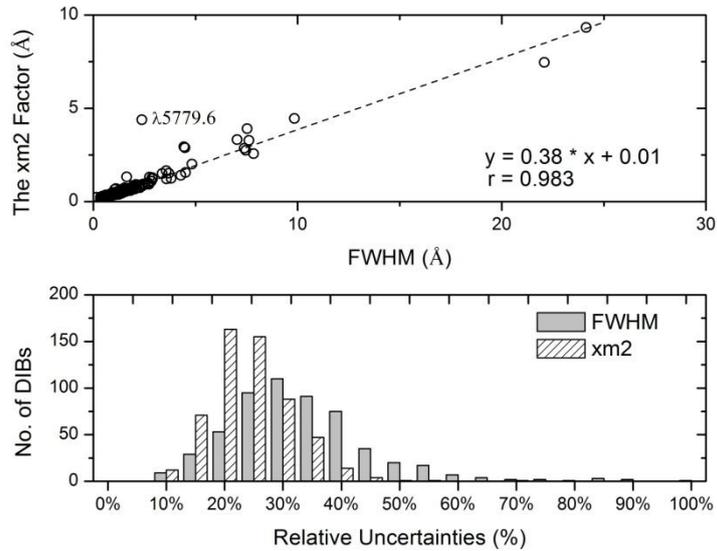

Figure 6. Comparison between FWHM and the xm2 factor, which is the apparent optical depth weighted variance of the DIB profile. Top panel: The two factors, both of which describe the width of the DIB profile, are well correlated by a linear model. The slope of the best-fit line is 0.38, which is close to the expected value for Gaussian profiles (0.42). Bottom panel: The xm2 factor tends to have smaller relative uncertainties than the FWHM. This is because the FWHM reported in this work depends on only three points in the spectrum, while the xm2 factor is derived from all points within the profile of each particular DIB. The two plots suggest that the xm2 factor in this work is well correlated with the FWHM but has smaller uncertainties, and may thus be a better way to describe the widths of DIB profiles.

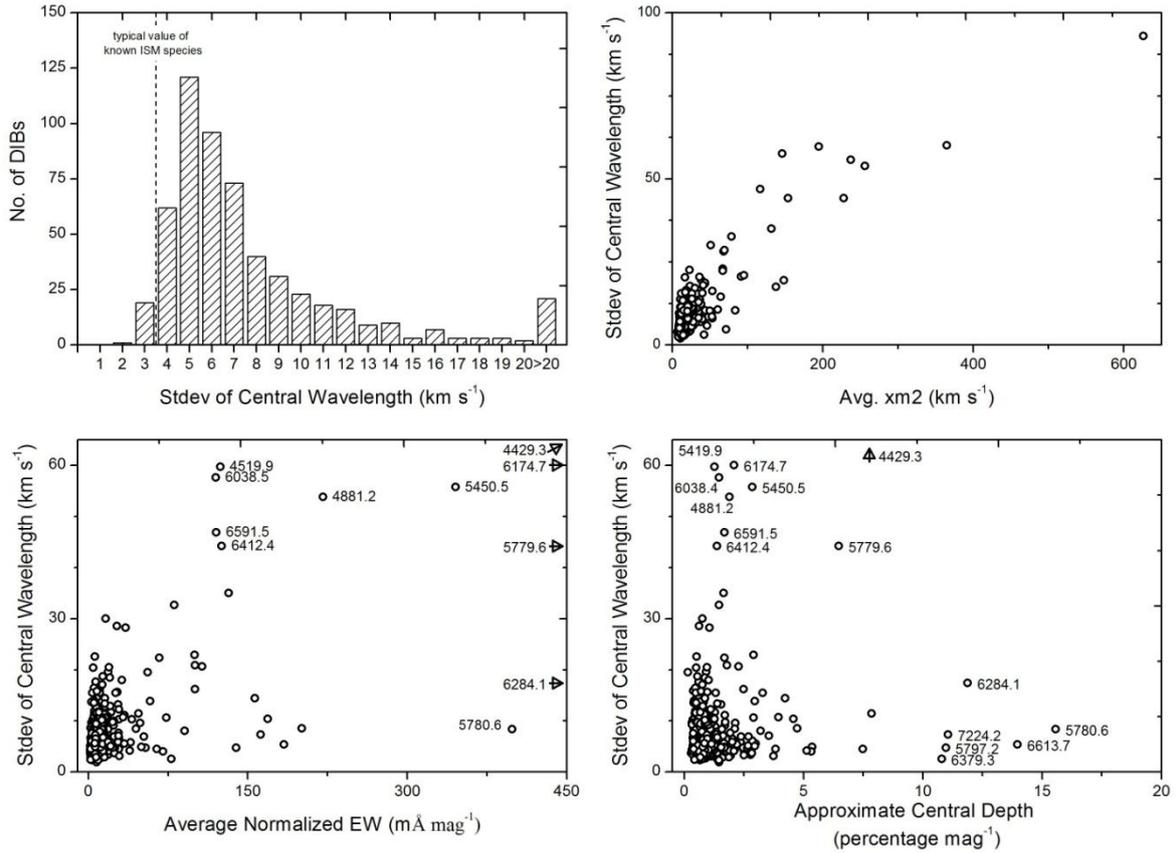

Figure 7. Top-left panel: Histogram of the standard deviation of central wavelengths for the 559 DIBs reported in Table 2. We found the average value of the standard deviation of DIB wavelengths to be 0.16Å, larger than the median value of six interstellar lines, which is 3.5 km s$^{-1}$ (noted by the dashed line). These interstellar lines are those of CH, CH$^+$, and K I, and are measured in the same spectra in the same way as the DIBs. Top-right panel: a linear correlation can be identified between the standard deviation of DIB wavelengths and the xm2 factor representing the width of the profile. Bottom-left panel: No obvious trend is identified when comparing to the average normalized EW of the DIB. Four DIBs outside this region are marked by arrows pointing to their location outside the plotting window (DIBs $\lambda\lambda$5779.6, 6174.6, and 6284.1 to the right, and DIB $\lambda$4429.3 to the upper right). Bottom-right panel: The fluctuation of central wavelength is compared to a surrogate central depth defined as the normalized EW divided by xm2. No trend can be identified but the high points are usually at the left side of the plot. These points are for DIBs with large width and medium strength. Note that these surrogates may differ from the actual central depths found in the spectrum.

# 7. SUMMARY

In this work, we selected spectra of twenty-five medium-to-highly reddened sight lines sampling a variety of ISM conditions and having background stars of different spectral types. A DIB detection survey was made using this data sample, aiming for a comprehensive catalog of DIBs between 4,000 and 9,000Å. Due to the distribution of stellar lines and telluric bands as well as other observational factors, we found the most sensitive wavelength region to be between ~5,500 and ~7,000Å. The resulting *Apache Point Observatory Catalog of Optical Diffuse Interstellar Bands* is based on -- and supersedes -- the Atlas Catalog of 545 DIBs toward HD 183143 and HD 204827 (Hobbs et al. 2008 and 2009), and on tests on detection percentages and $E_{B-V}$ correlation for final confirmation. About 85% of the 559 DIBs in the APO Catalog of DIBs have central wavelengths, widths and profiles that match those in the Atlas Catalog. Another ~10% of the DIBs in the APO Catalog of DIBs are updated from the Atlas Catalog, and 22 new DIBs are found in this work. Independent measurements were then made using a semi-automated program (*arcexam*) for DIBs not blocked or contaminated by telluric or stellar lines in the twenty-five program stars. By analyzing these measurements, we reach the following conclusions:

1) The detection percentages for individual DIBs peak around 50 - 60%, and 19 out of the 559 DIBs in the APO Catalog of DIBs have detection percentages greater than 90% in our data sample. Many of these frequently detected DIBs are strong and well studied. The high detection percentage makes them most ideal for correlation studies sampling a large number of sight lines. These DIBs are, in decreasing order of detection percentages as separated by semicolon, $\lambda\lambda$4963.9, 5779.6, 5780.6, 5797.2, 5849.8, 6203.6, 6284.1, and 6613.7 (with detection percentages of 100%); 4727.0, 6089.9, 6196.0, 6439.5, and 6660.7 (with detection percentages of 96%); 6376.1, 6379.3, 6449.3, and 6993.1 (with detection percentages of 92%).

2) The number of DIBs detected in a sight line is roughly proportional to the reddening, except for the heavily reddened sight lines toward VI Cyg 12 (due to the hard limit of 100% for the detection percentage). The linear relationship between $E_{B-V}$ and the number of DIBs detected holds for sight lines with $E_{B-V}$ less than ~ 2 mag.

3) We also found a weaker dependence for the number of DIBs detected in a given sight line on the spectral type of the background star. This is most likely due to the more severe stellar contamination for cooler stars, and is more clearly seen when the dependence on $E_{B-V}$ is removed. A lambda-shaped trend -- similar to those identified for several non-$C_2$ DIBs in Fan et al. (2017) -- can be identified between the detection percentage and the sight line $f_{H2}$, but this result is less secure due to the limited number of sight lines in this sample. Such a trend was identified for several non-$C_2$ DIBs in Fan et al. (2017). The possible appearance of this lambda-shaped trend in the overall detection percentage here is consistent with the $C_2$ DIBs (whose normalized EWs seem to increase with $f_{H2}$) being a small fraction of the total DIB population.

4) We compared the xm2 parameter reported by *arcexam*, which is the optical depth weighted variance of the DIB profile derived from all points between the integration limits, to the more commonly used width factor FWHM. The two parameters are linearly correlated, with xm2 ~ 0.38 * FWHM (close to the relationship expected for Gaussian profiles).

5) The typical standard deviation in the measured wavelengths of a given DIB is ~ 7.5 km s$^{-1}$ (average value), somewhat larger than the corresponding standard deviations for interstellar lines of CH, CH$^+$ or K I (which is ~ 3.5 km s$^{-1}$). The standard deviations of the DIB wavelengths go roughly as the square root of the DIB profile width (as expected), with no obvious dependence on EW or central depth.

6) Based on the new APO Catalog of DIBs, the difference in DIB populations between the two atlas sight lines (HD183143 and HD204827) is not as strong as previously thought. Of the 559 total DIBs in the APO Catalog, 416 have now been detected in both atlas sight lines -- though with significant differences in relative strengths. Relatively few new DIBs were found in this work, so that total number of DIBs is only slightly larger than the 545 identified in the two previous atlas papers (Hobbs et al. 2008 and 2009). We thus conclude that the populations of DIBs found in most sight lines in the solar neighborhood would lie within the list of DIBs in the *Apache Point Observatory Catalog of Optical Diffuse Interstellar Bands*.

We thank Adolf Witt and Ethan Polster for the special calculation of the mean radiation fields toward our stars. This effort was made possible by the creators of the Interstellar Radiation Field Calculator http://cads.iiap.res.in/tools/isrfCalc at the Indian Institute of Astrophysics, who combined the data from the HIPPARCOS catalog with Kurucz model atmospheres to allow these estimates to be made. We would like to express our gratitude to G. Galazutdinov for his very helpful and constructive comments as the referee of this work, and for sharing the high-resolution spectra with us. HF and GZ are supported by the National Natural Science Foundation of China under grant Nos. 11890694, 11603033, and 11703038. DGY was supported throughout the data acquisition parts of this project by NSF grant AST-1009603, DEW by NSF grant AST-1238926, JAD by NSF grant AST-1008424, TPS by NSF grant AST-1009929, BLR by NSF grant AST-1008801.

*Facilities*:   APO (ARCES)

**APPENDIX DIBS DETECTED IN EACH SIGHT LINE**

The DIB spectrum between 4000 and 8000Å is reconstructed using the measurements we made for each sight line. In this representation, Gaussian profiles are assumed for all detected DIBs where the central wavelength and FWHM are taken as the average values (column B and F of Table 2). The area beneath the Gaussian profile represents the EW of the DIB. This assumption is due to the lack of knowledge of the intrinsic DIB profiles, but we emphasize that the measurements were originally made by direct integration rather than profile fitting. The only exception is for DIB $\lambda$5779.6 whose FWHM is not a proper description of its width (Section 6.2). We use the general correlation between FWHM and the xm2 factor and estimate its FWHM to be 6.2Å.

The reconstructions of these spectra used ALL measurements of detected DIBs, including those contaminated by stellar or telluric lines, or when there is difficulties for the placement of continuum (although those measurements were omitted during the analysis made in previous sections). Thus, some of the DIBs might appear stronger/weaker than they really are when there is contamination from different sources. If measured, four absorption lines (CH band around 4300Å, $CH^+$ band around 4232Å, and K I doublet at 7665 and 7696Å) are included in the plot. We did not, however, include the other known atomic/molecular lines such as the strong Na I D lines around 5900Å.

The reconstructed spectra are arranged, from the top to the bottom, in the order of increasing $E_{B-V}$ and the plots are divided into four 1000Å-wide segments for better visualization. The star names are listed on the left end of each spectrum and can be easily associated with the information given in Table 1. The spectral types and $E_{B-V}$ values are provided in the first plot. The scale height for the plotting region of each spectrum is 10% of the continuum level. For some of the strongest DIBs in the most reddened sight lines, the profile might extend below 10% absorption. These parts of the spectra were truncated at 90% of the continuum and thus appear as a "flat bottom". This truncation is only to avoid crowding in the plots, and is not an indication that these DIBs are saturated, or an accurate representation of the true DIB profile.

The wavelength regions where telluric correction was preformed are yellow-shadowed. We use red arrows to indicate DIBs deemed unmeasurable in each sight line due interference from stellar or telluric lines, or difficulties in continuum placement. We cannot confirm if those DIBs are there. Broad DIBs are colored in red as their measurements may be less accurate given the limitations of the echelle spectra (Section 5.2). Finally, we use vertical lines to mark the central wavelengths of some of the strong and well-known DIBs (black), as well as the known interstellar species (red). These lines are labeled at the bottom by the wavelength for the DIBs, or by the name of the carrier for known ISM lines. The lines for known ISM lines are slightly shifted to avoid blocking their narrow profile.

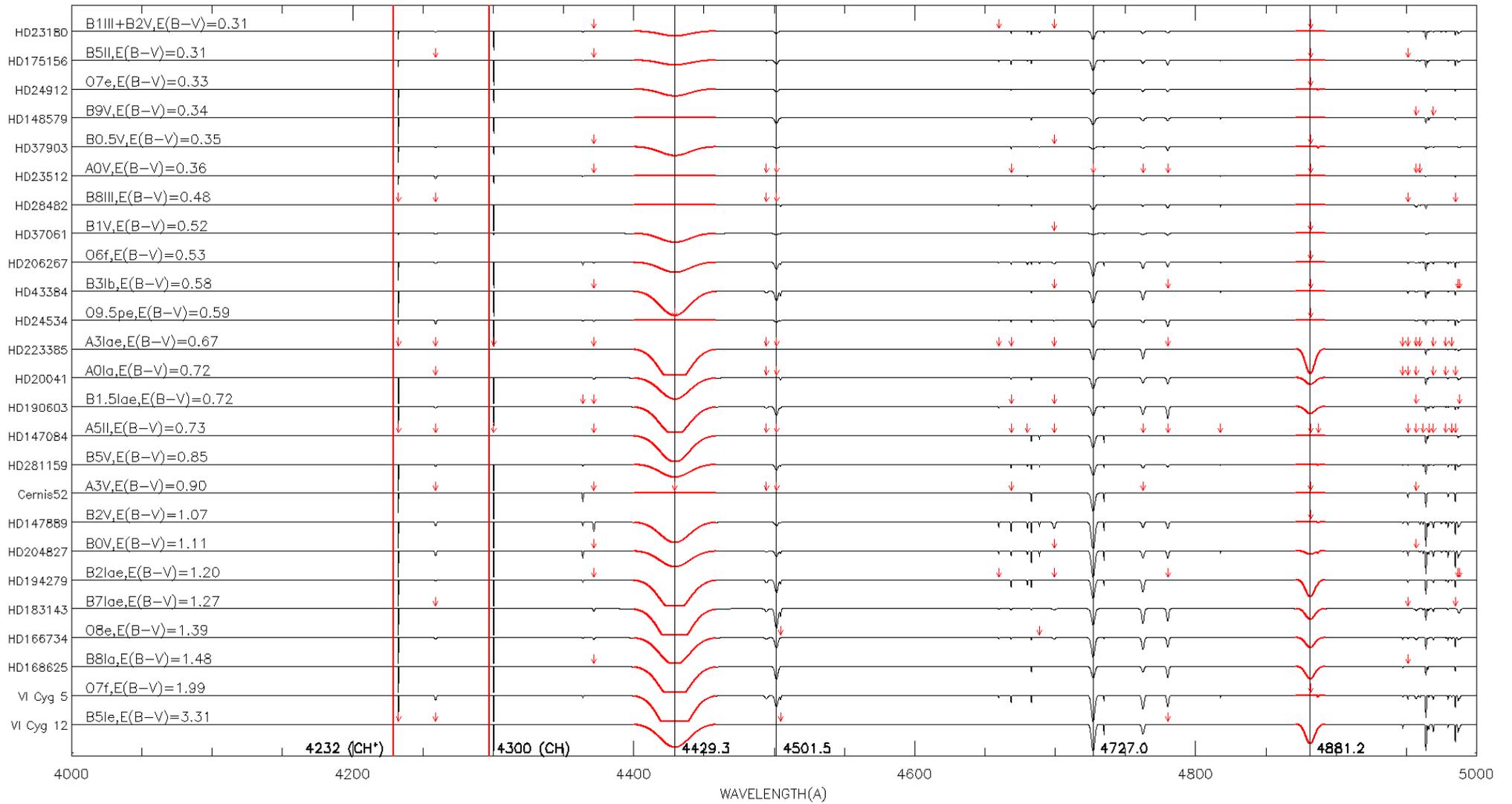

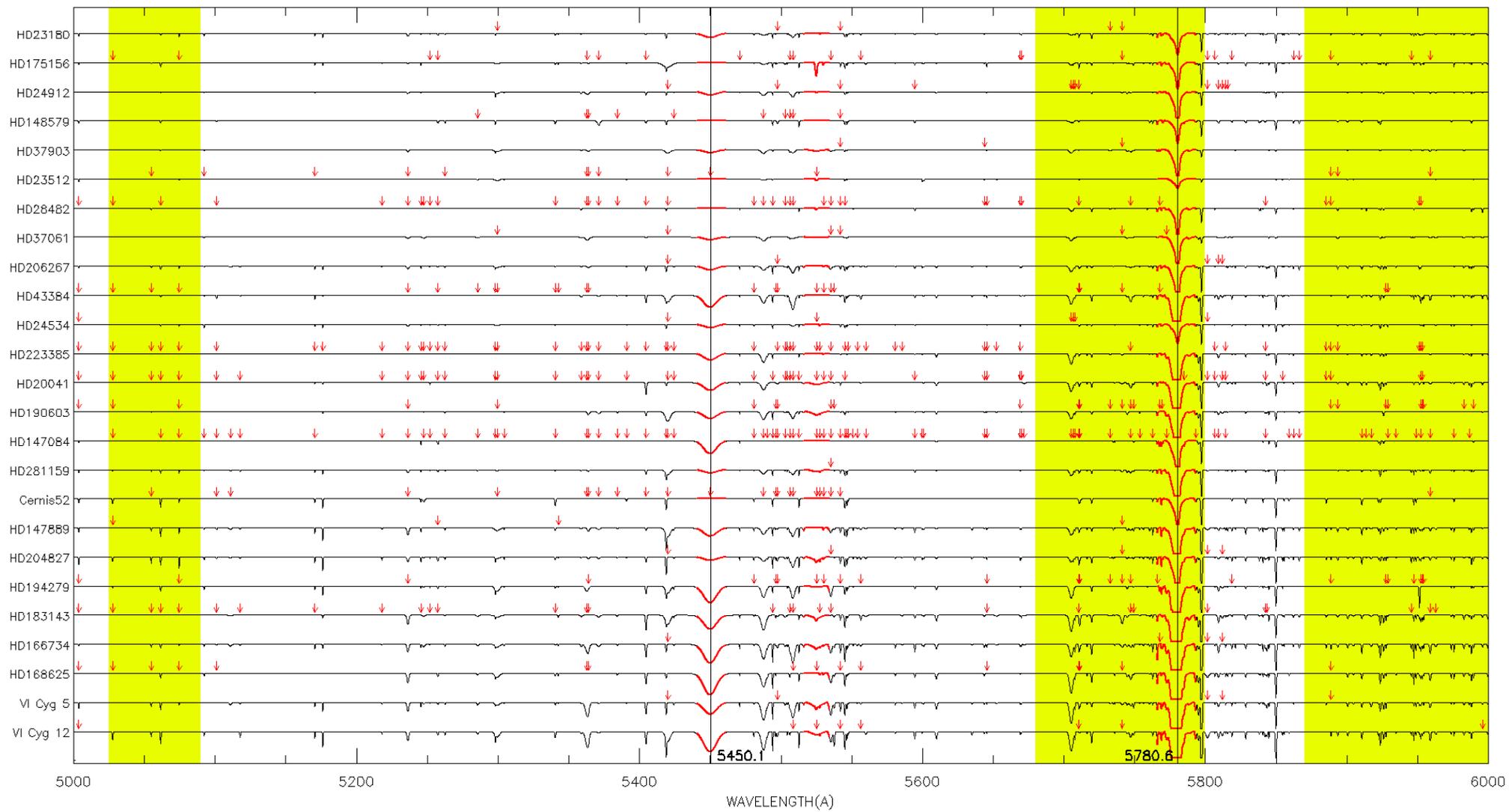

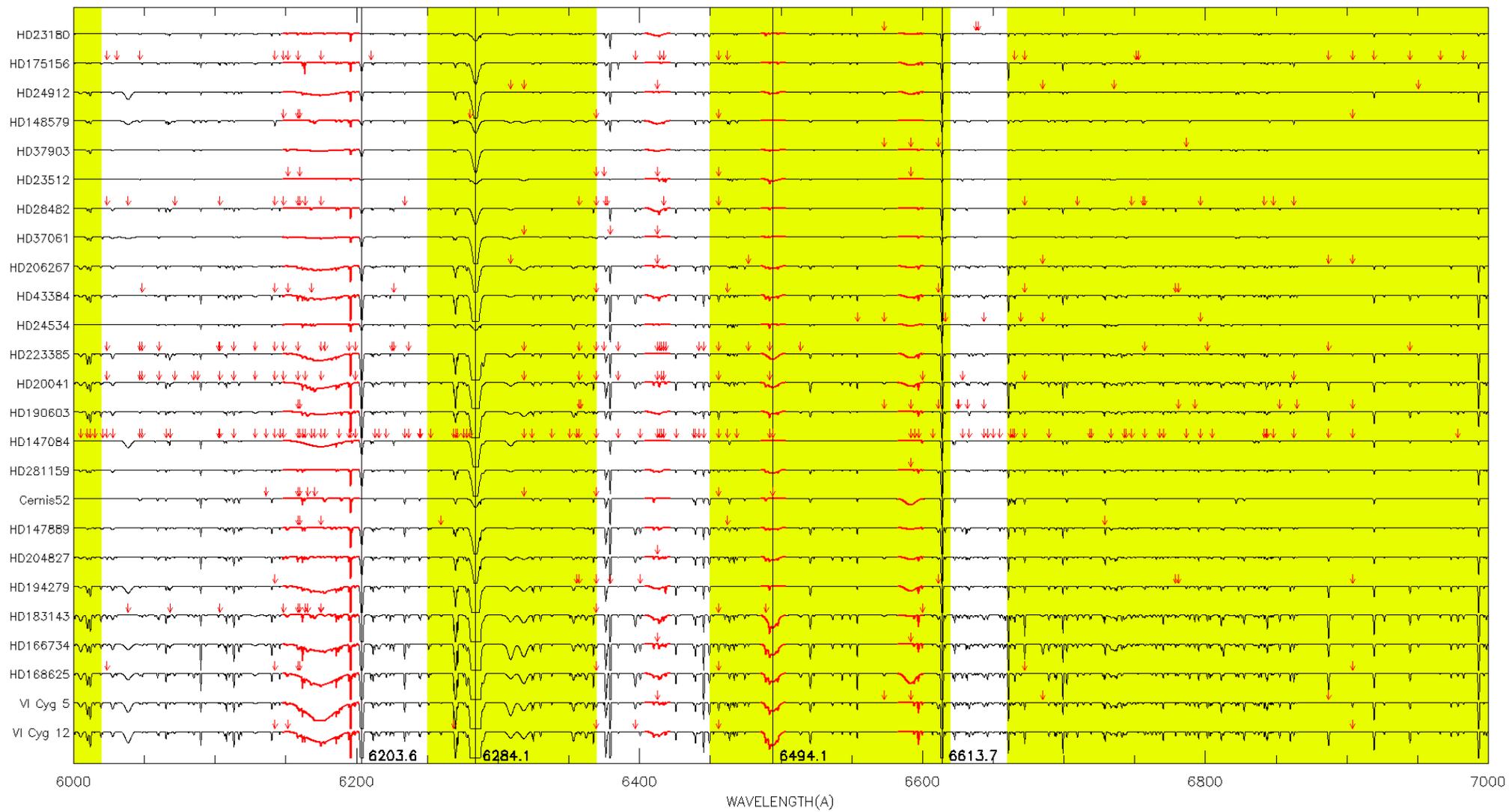

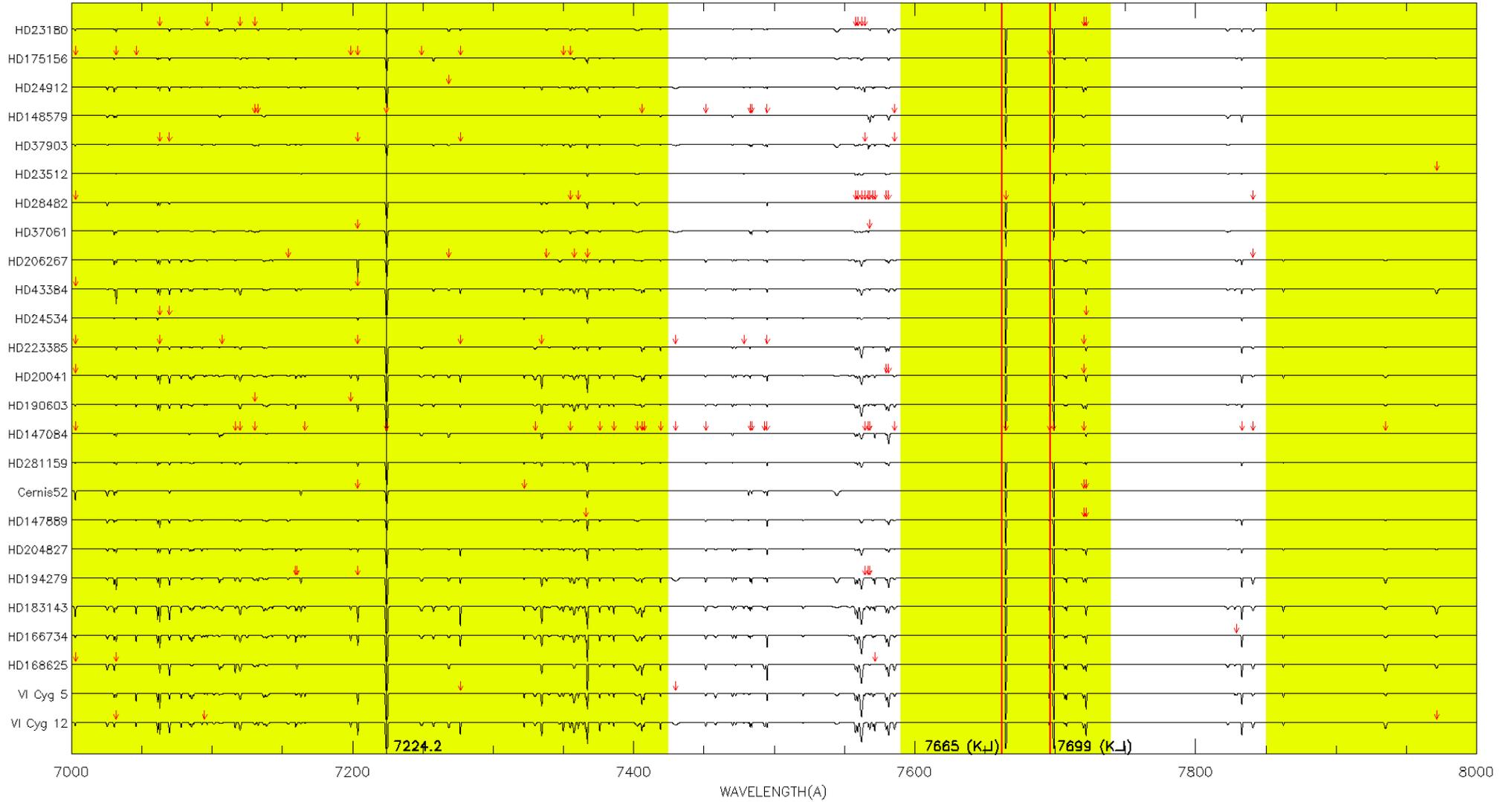

Figure A1. Synthetic spectra of the program sight lines in this work. These spectra were reconstructed assuming Gaussian profiles (which was not assumed during the measuring process), and arranged by increasing $E_{B-V}$. Telluric regions are shadowed. We use red arrows to mark DIBs deemed unmeasurable in each sightline. Profiles of broad DIBs are in red, and vertical lines mark strong DIBs and known ISM lines (slightly shifted).